%% file: mnras_main.tex
\documentclass[fleqn,usenatbib]{mnras}

\usepackage{newtxtext,newtxmath}

\usepackage[T1]{fontenc}
\usepackage{ae,aecompl}
\usepackage{booktabs}
\usepackage{soul}

\usepackage{enumitem}%
\usepackage{calc}
\usepackage{graphicx}	
\usepackage{amsmath}	
\usepackage{amssymb}	
\usepackage[dvipsnames]{xcolor}
\usepackage{xcolor}
\usepackage{makecell}
\usepackage{subcaption}
\usepackage{enumitem}
\usepackage{tipa}
\usepackage{accents}
\usepackage{amssymb}
\usepackage{amsfonts}

\definecolor{mypink}{rgb}{0.958, 0.188, 0.478}

\title[DanceCam]{DanceCam: atmospheric turbulence mitigation in wide-field astronomical images with short-exposure video streams}

\author[Bialek et al.]{
\parbox{\textwidth}{
Spencer Bialek$^{1,2}$\thanks{E-mail: sbialek@uvic.ca.  
},
Emmanuel Bertin$^{2,3}$,
S{\'e}bastien Fabbro$^{1,4}$,
Herv{\'e} Bouy$^{5, 6}$,
Jean-Pierre Rivet$^{7}$,
Olivier Lai$^{7}$,
Jean-Charles Cuillandre$^{3}$
\\
}
\\
$^{1}$Department of Physics and Astronomy, University of Victoria, Victoria, BC, V8W 3P2, Canada\\
$^{2}$Canada–France–Hawaii Telescope, Kamuela, HI 96743, USA \\
$^{3}$AIM, CEA, CNRS, Université Paris-Saclay, Université Paris Cité, F-91191 Gif-sur-Yvette, France\\
$^{4}$National Research Council Herzberg Astronomy and Astrophysics, Victoria, BC, Canada\\
$^{5}$Laboratoire d’Astrophysique de Bordeaux, CNRS and Université de Bordeaux, Allée Geoffroy St. Hilaire, 33165 Pessac, France\\
$^{6}$Institut Universitaire de France \\
$^{7}$Universit\'e C\^ote d'Azur, Observatoire de la C\^ote d'Azur, CNRS, Laboratoire J.--L. Lagrange, F-06304 Nice Cedex 4, France\\
}

\date{Accepted 2024 April 05. Received 2024 March 16}

\pubyear{2024}

\begin{document}
\label{firstpage}
\pagerange{\pageref{firstpage}--\pageref{lastpage}}
\maketitle



\input{Sections/Abstract/abstract}
\input{Sections/Introduction/intro}
\input{Sections/Methods/methods}
\input{Sections/Data/data}
\input{Sections/Results/results}

\input{Sections/Discussion/discussion}
\input{Sections/Conclusions/conclusions}



\section*{Acknowledgements}


We acknowledge and respect the l\textschwa\textvbaraccent {k}$^{\rm w}$\textschwa\ng{}\textschwa n peoples on whose traditional territory the University of Victoria stands and the Songhees, Esquimalt and $\underaccent{\bar}{\rm W}$S\'ANE\'C peoples whose historical relationships with the land continue to this day.

This research has received funding from the European Research Council (ERC) under the European Union’s Horizon 2020 research and innovation programme (grant agreement No 682903, P.I. H. Bouy), from the French State in the framework of the "Investments for the future" Program, IdEx Bordeaux, reference ANR-10-IDEX-03-02, and from the French "Projets nationaux" PNCG, PNHE and PNP.

We thank the National Sciences and Engineering Research Council of Canada for funding through the Discovery Grants and CREATE programs, and the Canada-France-Hawaii Telescope -- where most of this work was done -- for support both financially and from the excellent staff.
We also thank Prof.\,Kim A.\,Venn for helpful comments and support on this project.
This research was enabled in part by support provided by Calcul Québec (\href{https://www.calculquebec.ca/en/}{calculquebec.ca}) and the Digital Research Alliance of Canada (\href{http://alliancecan.ca}{alliancecan.ca}).

\section*{Code availability}

A public repository of the code which produced the figures in this paper is available at \href{https://github.com/DanceCam/dancelibpublic}{https://github.com/DanceCam/dancelibpublic}. The rest of the code will be shared on reasonable request to the corresponding author.

\section*{Data availability}

Videos of the simulations and the observations, together with additional information about the DanceCam project are available at \href{https://dancecam.info}{https://dancecam.info}. All other data underlying this article will be shared on reasonable request to the corresponding author. 



\bibliographystyle{mnras}
\bibliography{mnras_main.bib}



\input{Sections/Appendix/appendix}
\bsp	
\label{lastpage}
\end{document}

%% file: Sections/Abstract/abstract.tex
\begin{abstract}
We introduce a novel technique to mitigate the adverse effects of atmospheric turbulence on astronomical imaging. Utilizing a video-to-image neural network trained on simulated data, our method processes a sliding sequence of short-exposure ($\sim$0.2s) stellar field images to reconstruct an image devoid of both turbulence and noise. We demonstrate the method with simulated and observed stellar fields, and show that the brief exposure sequence allows the network to accurately associate speckles to their originating stars and effectively disentangle light from adjacent sources across a range of seeing conditions, all while preserving flux to a lower signal-to-noise ratio than an average stack. This approach results in a marked improvement in angular resolution without compromising the astrometric stability of the final image.

\end{abstract}

\begin{keywords}
methods: data analysis
methods: observational
techniques: image processing
atmospheric effects
\end{keywords}

%% file: Sections/Introduction/intro.tex
\section{Introduction}

Modern astronomy demands data of unparalleled precision and resolution to further its insights and discoveries. In their pursuit of this goal, astronomers often wrestle with the challenges imposed by the Earth's atmosphere. Due to random fluctuations in pressure, temperature, and wind speed, turbulence arises in the atmosphere along with a stratified refractive index distribution. The rapidly varying nature of these fluctuations causes the wavefront of incoming light to be dynamically distorted, leading to the twinkle of starlight. Short exposure images reveal the starlight to be composed of speckles with randomly varying shapes and positions around a central point. As a result, long exposures lead to the degradation of image resolution and the overall effect can be framed as a blurring operation from a bell-like Point Spread Function (PSF). Atmospheric turbulence thus causes a significant loss in the spatial resolution of astronomical images and is a major impediment to obtaining high-quality imaging data \citep{roddier1981v}. 

The only way for a ground-based telescope to combat turbulence is by recording its effects in real-time and using the information from these snapshots to move the speckles back to a central point. Adaptive optics (AO) is one such method that has emerged as a revolutionary tool to counteract these disturbances, enabling telescopes to achieve near diffraction-limited observations \citep{beckers1993adaptive, hardy1998adaptive}. AO systems include a wavefront sensor to measure the wavefront distortions, from a laser or natural guide star, caused by the atmosphere, and then compensate for these distortions, typically with a deformable mirror, on millisecond timescales. The performance of AO has seen remarkable improvements over time, with modern systems capable of delivering images with resolutions that rival those from space telescopes \citep{davies2012adaptive}. 

However, AO is not without its limitations. Traditional AO systems often have a limited field of view, correcting only a small region around the guide star. Their effectiveness is also contingent on the presence of a \textit{bright} guide star or the use of artificial laser guide stars, limiting where and when they can be used. To address these limitations, modern AO systems have evolved to incorporate advanced techniques such as Multi-Conjugate Adaptive Optics (MCAO) and Ground-Layer Adaptive Optics (GLAO). MCAO uses multiple deformable mirrors conjugated to specific layers of the atmosphere to correct over a wider field of view, while GLAO aims to provide a uniform correction over a wide field by primarily correcting the turbulence close to the ground \citep{rigaut2000principles, johnston1994analysis, tokovinin2004seeing}. While these systems have achieved impressive corrections over wider fields \citep[e.g.][]{massari2016astrometry, abdurrahman2018improved}{}{}, the field of view still remains significantly smaller compared to what is achievable with uncorrected modern imaging fields, and there are additional challenges related to the cost and complexity of implementing and maintaining advanced AO systems. Time overheads during observations are also often an issue.

An alternative method that has gained traction in the astronomical community is high cadence imaging, in particular popularized with the ``lucky imaging" technique. Lucky imaging capitalizes on the intermittency of turbulence and the brief moments of atmospheric stability. By taking a rapid series of short-exposure images, only the sharpest frames -- those taken during moments of optimal seeing conditions -- are selected. These ``lucky" frames are then aligned and combined to produce a single high-resolution image. While lucky imaging can achieve impressive resolutions over wide fields \citep[][]{mackay2018gravitycam}{}{}, especially for brighter targets, its effectiveness is inherently tied to the whims of the atmosphere. In poor seeing conditions, the probability of capturing lucky frames decreases, making the technique less effective \citep[][]{faedi2013lucky}{}{}. 

Another popular seeing mitigation technique using high cadence imaging is fast guiding, which focuses on compensating the image jittering caused by atmospheric turbulence. This is only effective in the regime where the tilt component of the wavefront distortions outweighs the sum of all other contributions to the total phase error budget, that is, when the PSF features a dominant speckle which can be used to track and compensate for random image motions on the focal plane. In practice, the correction is most effective for telescopes with apertures $\approx 3-4$ times the Fried parameter \citep{fried1966optical,young1974}. The compensation may be done using, e.g., a tip-tilt mirror \citep[e.g.,][]{mcclure1989}, two-dimensional synchronous charge transfers \citep{tonry1997}, or shift-and-add stacking of short exposures \citep{bates1980}. In the regime where the PSF features several speckles, and where a sufficiently bright star is available inside the region of interest, one may also consider holographic image reconstruction \citep{schodel2013holographic}.

In all cases of fast guiding, the correction remains only effective over a limited solid angle around the reference star(s) (isokinetic or isoplanatic disk, depending on the type of correction). The diameter of the disk depends on the altitude of the turbulent layer and ranges from a degree or more for the ground layer down to a few arcseconds for the highest layers. Hence in practice this type of method is only applicable to small patches individually, and correcting apparent motions over a wide field-of-view requires adjusting a ``rubber'' focal plane model \citep{kaiser2000}, which consists of a distorted virtual pixel grid whose deformations are continuously controlled by a number of guide stars over its surface.

As the field of astronomical instrumentation continues to push the boundaries of observational capabilities, there is a pressing need for more versatile and cost-effective solutions that can operate under a broader range of conditions. With the rapidly evolving landscape of digital technology, there is a growing opportunity for machine learning (ML) methods to meet these demands and help fill in some of the gaps of AO and lucky imaging systems.

ML has already been successfully employed to mitigate atmospheric turbulence effects in long-range imaging applications \citep[][]{nieuwenhuizen2019deep, vint2020analysis, hoffmire2021deep, zhang2022imaging}. These methodologies primarily rely on training ML algorithms on a collection of artificially distorted images, enabling them to predict the known ground truth and rectify turbulence-induced aberrations, though there are other methods which attempt to accomplish this in an unsupervised way \citep[e.g.,][]{li2021unsupervised}{}{}. In essence, ML can be used as a digital counterpart to lucky imaging, where instead of relying on elusive high quality exposures, algorithms \textit{infer} the turbulence-free image from a video sequence of short exposures. This ``DanceCam'' approach, as we call it, offers the potential to consistently produce sharp, high-resolution images, even in less-than-ideal seeing conditions and with a more efficient use of telescope time. 

Despite the success in long-range imaging, the adaptation and application of ML techniques to turbulence in astronomical images remains largely uncharted, presenting a promising frontier for future research. The uniqueness of astronomical imaging poses specific challenges and requirements that differentiate it from long-range imaging. Characteristics like the extreme range of object brightness, object sizes and scales, and the requirement for ultra-high resolution and precision measurements, make astronomical imaging a domain where a distinct approach is required \citep{tyson2022principles}.

In this paper, we present a novel method to combat the deleterious effects of atmospheric turbulence in astronomical images using ML trained on simulations of turbulent and noisy video-streams of stellar fields. Section \ref{section:methods} outlines the method used to simulate atmospheric turbulence and the ML methods used in this study. Section \ref{section:data} describes both how the simulated datasets were created and how real test data was collected from the C2PU telescope. Section \ref{section:results} is an overview of the main results from evaluating the proposed method on simulated and real data. Section \ref{section:discussion} summarizes the strengths and limitations of the proposed method, and concluding remarks are in Section \ref{section:conclusions}.

%% file: Sections/Methods/methods.tex
\section{Methods}
\label{section:methods}

\subsection{Simulating atmospheric turbulence}

We implement a variation of the ``split-step" method \citep[e.g.][]{chatterjee2014split}{}{} for simulating the propagation of a wavefront through the atmosphere, in which the atmosphere is decomposed into several distinct layers which perturb the wavefront as it passes through. This section describes the theoretical framework used for the split-step simulations.

The Kolmogorov theory of turbulence \citep[][]{kolmogorov1941equations, kolmogorov1941local}{}{} was one of the first models used for describing the statistical properties of atmospheric turbulence \citep[][]{frisch1995turbulence}{}{}. It assumes that the turbulence is isotropic, homogeneous, and fully developed, which means that the turbulence has reached a statistically steady state. The Kolmogorov model is characterized by a power-law scaling of the spatial frequency spectrum of the turbulence, and the refractive-index power spectral density is given by:

\begin{equation*}
\Phi_n = 0.033~C_n^2~k^{-11/3} ~~~\\
\textrm{ for } \frac{1}{L_0} \ll k \ll \frac{1}{l_0},
\end{equation*}
where $L_0$ is the outer scale, i.e. the average size of the largest eddies, $l_0$ is the inner scale, i.e. the average size of the smallest eddies, $C_n^2$ is the refractive index structure constant, which is a measure of the strength of the turbulence, and $k$ is the angular spatial frequency. The spatial frequency spectrum in the Kolmogorov model has a power-law behavior with an exponent of $-11/3$, which means that high spatial frequencies are strongly attenuated by atmospheric turbulence.

There exist more sophisticated models which include inner-scale and outer-scale factors to improve fits between theory and experiment. In the modified von K{\'a}rm{\'a}n model, for example, the refractive-index power spectral density is given by:

\begin{equation*}
\Phi_n = 0.033~C_n^2~\frac{\exp{(-k^2/k_m^2)}}{(k^2 + k_0^2)^{11/6}},
\end{equation*}
where $k_m = 5.92/l_0$ and $k_0 = 2\pi/L_0$. The modified von K{\'a}rm{\'a}n model is used in the atmospheric turbulence simulations in this study because it provides a more realistic description of the statistical properties of the turbulence than the simpler Kolmogorov model, at the expense of more free parameters.

\begin{figure}
\centering
\includegraphics[width=0.3\textwidth]{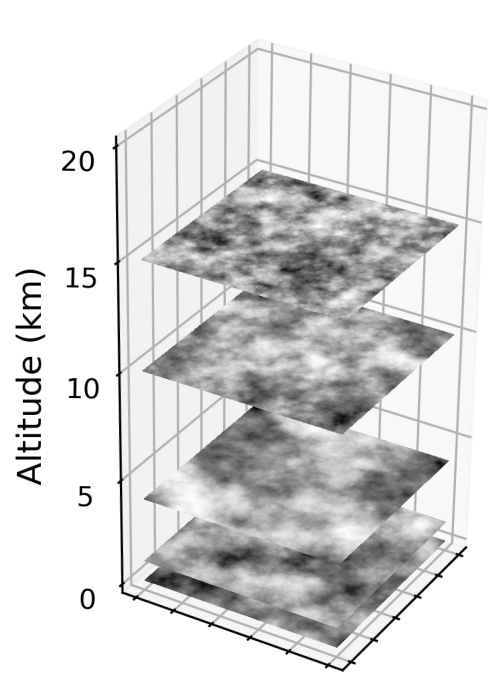}
\caption{An example of the phase screens used in the simulation pipeline. For every simulated video sequence, each layer is initialized with a different $r_0$ and wind speed to account for varying seeing conditions.}
\label{fig:phase-screen-layers}
\end{figure}

\begin{figure}
    \centering
    \begin{subfigure}{.6\linewidth}
        \centering
        \includegraphics[width=\linewidth]{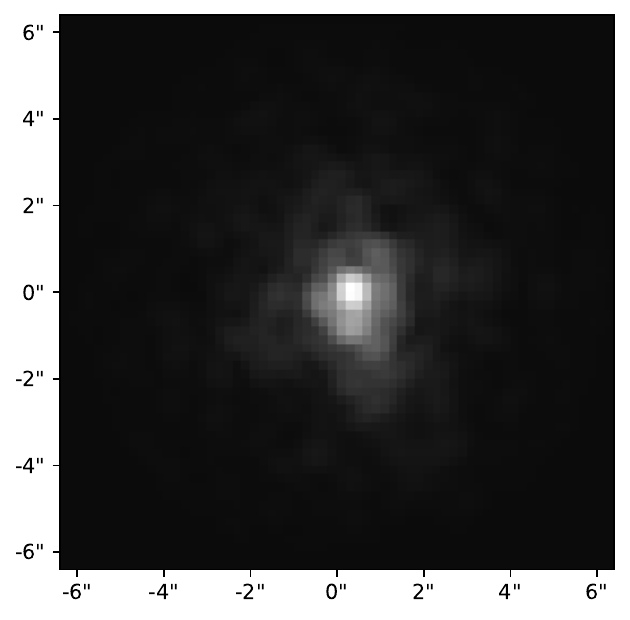}
        \caption{}
        \label{fig:1m_good_seeing_PSF}
    \end{subfigure}%
    \hfill
    \begin{subfigure}{.6\linewidth}
        \centering
        \includegraphics[width=\linewidth]{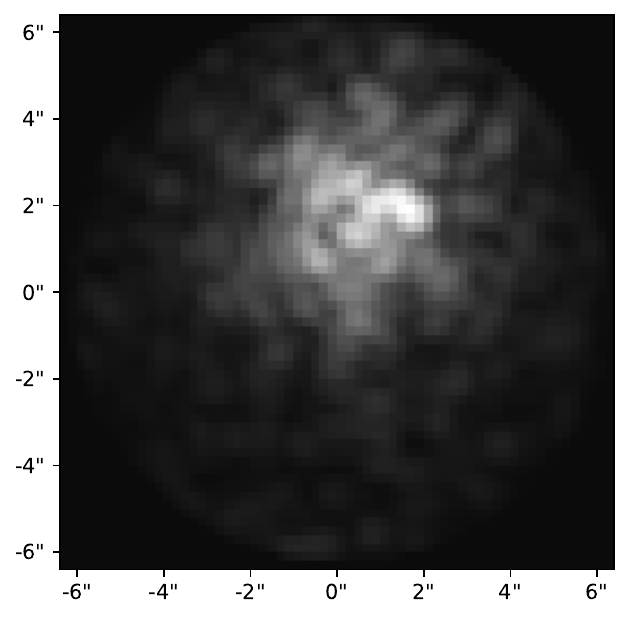}
        \caption{}
        \label{fig:1m_bad_seeing_PSF}
    \end{subfigure}
    \caption{The PSFs of the same star imaged with a 1m telescope in (a) mild turbulence (D/$r_0 \approx 3$)  and (b) strong turbulence (D/$r_0 \approx 7$). In the case of mild turbulence, the PSF is mostly concentrated in the centre with a clear Airy pattern around it, whereas with the strong turbulence, several Fried parameter length-scales can fit within the area of the telescope's aperture and so the light is distributed away from the centre in multiple ``speckles".}
    \label{fig:PSF_examples}
\end{figure}

Atmospheric turbulence is usually modeled using phase screens. A phase screen is a realization of the two-dimensional random phase distortion field introduced by turbulence at a particular altitude. In the modified von K{\'a}rm{\'a}n model, the statistical properties of the phase power spectral density are given by\footnote{See, e.g., \cite{schmidt2010numerical} for the full derivation.}:

\begin{equation}
\Phi_{\phi} = 0.023~r_0^{-5/3}~(f + 1/L_0^2)^{-11/6} \exp{(-1.126~l_0^2~f)},
\label{eq:mvK_phasePSD}
\end{equation}
where $f$ is the frequency in cycles/m, and $r_0$ is the Fried parameter, which represents the size of the region over which the wavefront distortion due to atmospheric turbulence is roughly constant. $r_0$, $L_0$, and $l_0$ may differ from layer to layer. Figure \ref{fig:phase-screen-layers} shows an example of the phase screens computed in our simulation software.

To calculate the impact of the phase screens on the wavefront of the light, we leverage the mathematical framework of Fourier optics in the Fresnel regime to describe the propagation of light through a series of planes on the line of sight, starting from the highest altitude layer.

In the case of atmospheric turbulence, the $i$th layer carries a phase screen represented by the two-dimensional phase distribution $\phi_i(x,y)$, where $x$ and $y$ are the spatial coordinates on the plane.
This phase screen interacts with a monochromatic incident wavefront, characterized by a complex amplitude $W_i(x,y)$, to generate the amplitude of the emergent wavefront
\begin{equation}
\label{eq:applyphasescreen}
\tilde{W}_i(x, y) = W_i(x,y) \exp j\phi_i(x,y).
\end{equation}

In the Fresnel (paraxial) approximation, the distorted wavefront is propagated to the next layer at distance $d = | z_{i+1} - z_i |$ by convolving with the impulse response of free-space propagation:
\begin{equation}
\label{eq:fresnel}
W_{i+1}(x,y) = h_i(x,y, |z_i - z_{i+1}|) * \tilde{W}_i(x,y),
\end{equation}
where
\begin{equation*}
h(x, y, d) = \frac{e^{2j\pi\frac{d}{\lambda}}}{j\lambda d} \exp j\pi \frac{x^2 + y^2}{\lambda d},
\end{equation*}
$z_i$ the altitude of the $i$th layer and $\lambda$ the wavelength.
(\ref{eq:fresnel}) is more conveniently computed in its ``angular spectrum'' form in Fourier space \citep{schmidt2010numerical}:
\begin{equation}
\mathcal{F}W_{i+1}(f_x,f_y) = H_i(f_x, f_y)\mathcal{F}\tilde{W}_i(f_x, f_y),
\end{equation}
with
\begin{equation*}
H(f_x, f_y, d) = e^{2j\pi\frac{d}{\lambda}} e^{-j\pi\lambda d (f_x^2 + f_y^2)},
\end{equation*}
where $f_x$ and $f_y$ are the spatial frequencies on the plane.

To mitigate aliasing effects in Fresnel propagation, we apodize the incident planewave from the distant point source using a cosine-tapered (Tukey) radial window \citep[e.g.,][]{harris1978windows}, with inner and outer diameters 1.5$D$ and 2$D$, respectively, where $D$ is the diameter of the telescope entrance pupil.

After cascading through $n$ atmospheric layers, the simulated wavefront reaches the telescope where the formalism remains the same except that pupil functions replace $\exp j\phi_i(x,y)$ in (\ref{eq:applyphasescreen}), first with a circular obstruction with diameter $D'$ by the secondary mirror or prime focus:
$$ P_{n+1}(x,y) = \begin{cases} 1, & \text{if } x^2 + y^2 \geq \left(\frac{D'}{2}\right)^2 \\ 0, & \text{otherwise} \end{cases}, $$
and finally by the circular windowing from the primary mirror.

$$ P_{n+2}(x,y) = \begin{cases} 1, & \text{if } x^2 + y^2 \leq \left(\frac{D}{2}\right)^2 \\ 0, & \text{otherwise} \end{cases} $$

\begin{figure*}
    \centering
    \includegraphics[width=1\textwidth]{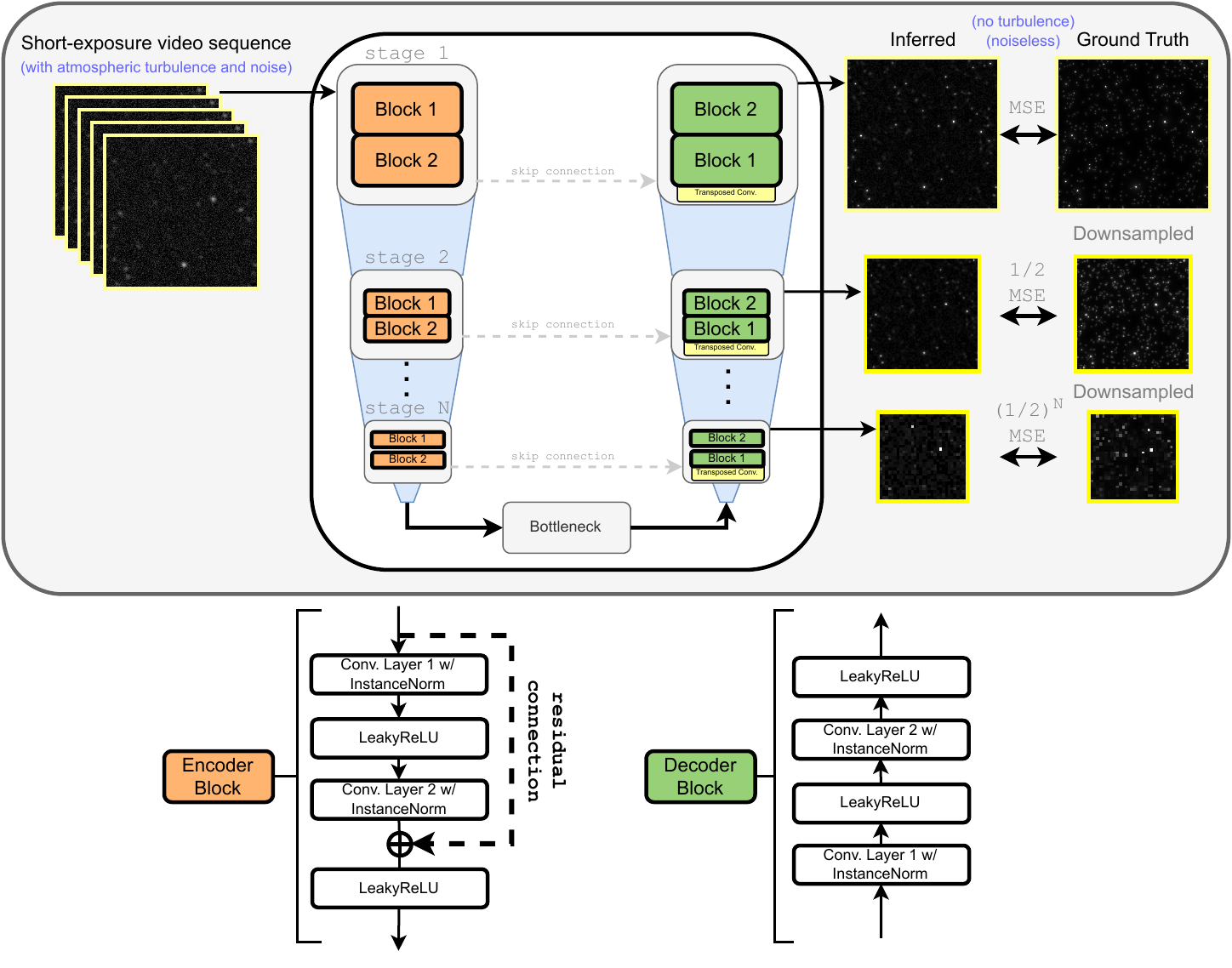}
    \caption{The DanceCam Residual U-Net architecture. A set of simulated short-exposure video streams of stellar fields -- with turbulence and noise -- along with their corresponding ground truth frames -- with no turbulence or noise -- is used to train the model. Instead of a single output, the model additionally has outputs from each stage in the decoder which are compared to downsampled versions of the ground truth using a weighted mean-squared error (MSE) loss function. Once trained, either a simulated or real video stream can be used as input and only a single (not downsampled) inferred image is retrieved.} 
    \label{fig:unet}
\end{figure*}

Assuming perfect optics, the (incoherent) optical Point Spread Function on the focal plane is obtained by taking the square modulus of the inverse Fourier transform of the complex wavefront amplitude on the entrance pupil:

$$ \mathrm{PSF}(x,y) = \left| \int_{-\infty}^{\infty} \int_{-\infty}^{\infty} \tilde{W}_{n+2}(x',y') e^{2 j \pi \left(\frac{xx' + yy'}{\lambda F}\right)} dx' dy' \right|^2,$$
where $F$ is the effective focal length of the instrument.
The PSF describes the response of the telescope and atmosphere to a point source object, and can be used to estimate the resolution of the imaging system.
Finally, the PSF is convolved with the Intra-Pixel Response Function, which we assume to be a perfect two-dimensional boxcar the size of the camera pixel.
Figure \ref{fig:PSF_examples} shows what the PSF of a star looks like when imaged with a 1m aperture telescope through an atmosphere with different strengths of turbulence (characterized by $r_0$ in Equation \ref{eq:mvK_phasePSD}). 

The images are sampled at the camera pixel resolution, and initially generated noiseless.
A spatially constant sky background is added and a constant gain factor is applied. Finally, realizations of noisy images are generated following a Poisson-Gaussian mixture representing the shot noise of the collecting instrument plus stationary white noise from the readout electronics.

The entire simulation pipeline is written with PyTorch so that GPUs could be maximally utilized with Fast Fourier Transforms \citep{brigham1967fast}. This results in the capability to render $\sim$150,000 PSFs per second, which is a couple orders of magnitude faster than other similar implementations \citep[e.g.][]{hardie2017simulation}.
Maximizing efficiency in the simulation pipeline is required because the datasets we generate contain hundreds of thousands of frames.

\subsection{Deep learning inference of turbulence-free images}

\subsubsection{Model architecture}

\begin{figure*}
    \centering
    \begin{subfigure}{1\linewidth}
        \centering
        \includegraphics[width=\linewidth]{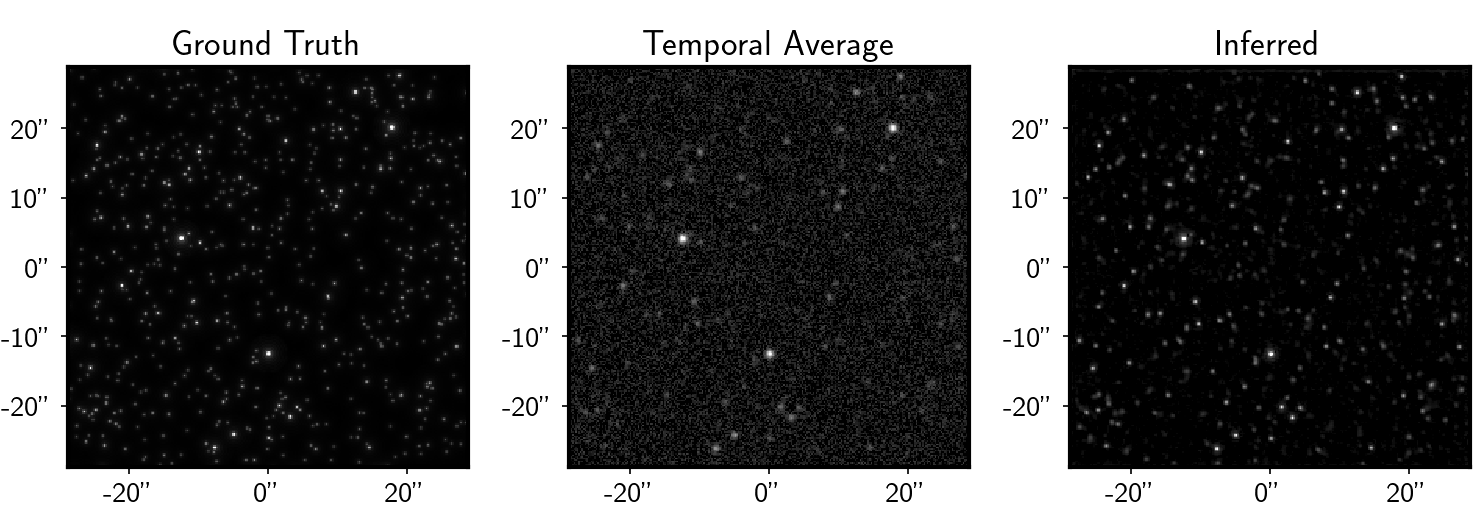}
        \caption{}
        \label{fig:1m_good_seeing_gt.avg.infer}
    \end{subfigure}%
    \hfill
    \begin{subfigure}{1\linewidth}
        \centering
        \includegraphics[width=\linewidth]{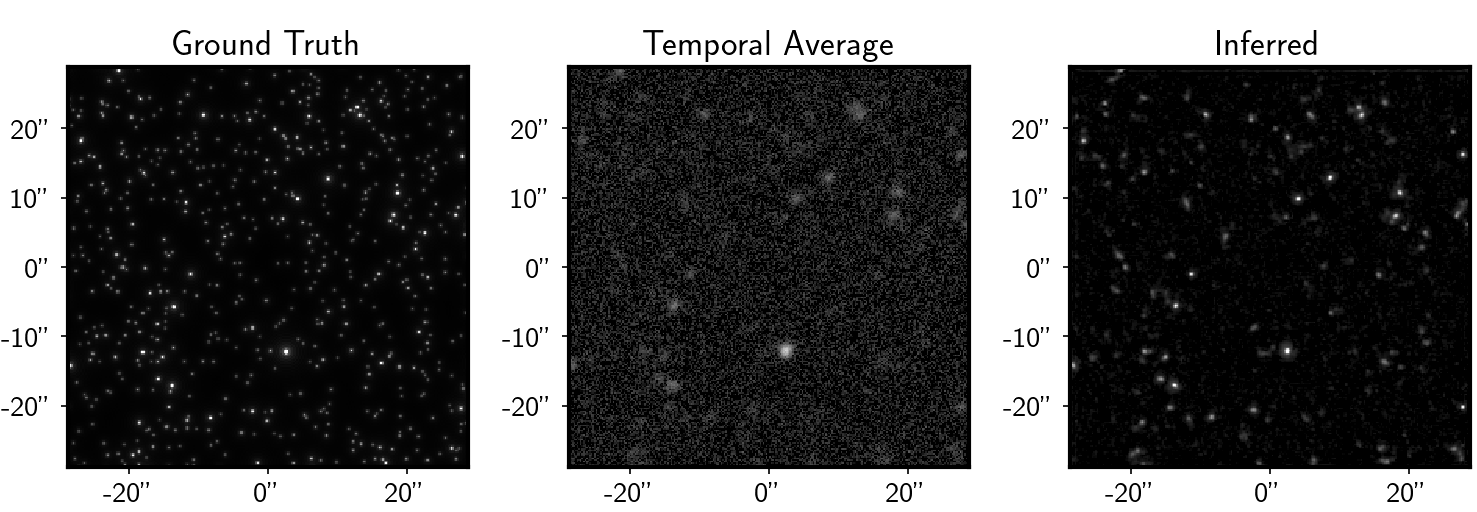}
        \caption{}
        \label{fig:1m_bad_seeing_gt.avg.infer}
    \end{subfigure}
    \caption{Two examples highlighting the ability of the proposed method to remove the effects of atmospheric turbulence and produce a sharp, clear image. 6-second sequences of random stellar fields were simulated with (a) 0.7" seeing and (b) 1.4" seeing, and the ground truth, temporally averaged sequence, and inferred frames are shown here. 
    }
    \label{fig:examples_inferred_goodbadseeing}
\end{figure*}

The cornerstone of our proposed method is the application of the Residual U-Net, a variant of the traditional U-Net architecture known for its proficiency in semantic segmentation and image reconstruction tasks \citep[][]{ronneberger2015u, cciccek20163d, yao2018pixel, zhang2018road, mizusawa2021computed}{}{}. The model used in this study, developed by the Medical Imaging Computing Group at the German Cancer Research Center (MIC-DKFZ)\footnote{\url{https://github.com/MIC-DKFZ/dynamic-network-architectures}}, is part of a widely-used Python package used for deep learning-based biomedical image segmentation \citep[][]{isensee2021nnu}{}{}. It was selected based on its unique architecture and inherent properties -- detailed below -- that make it suitable for our image reconstruction task.

A U-Net can be defined as a combination of a contracting path (encoder) and an expansive path (decoder), bridged by a bottleneck which helps to reduce the computational complexity of the model. The encoder performs consecutive convolutions and downsampling to output \textit{feature maps}, learning the contextual information while decreasing the spatial dimension of the input. The expansive path, on the other hand, performs transposed convolutions on the feature maps and then concatenates them with the corresponding feature maps from the encoder (skip connections), allowing the decoder to merge high-level features with preserved local information. 

The U-Net is composed of stages, each containing a certain number of \textit{blocks}. Each block includes convolutional layers with activation and normalization functions -- in our case, the leaky variant of the Rectified Linear Unit (\texttt{LeakyReLU}) activation and instance normalization (\texttt{InstanceNorm}) were used, noting that \texttt{InstanceNorm} was chosen for its ability to perform well on smaller batch sizes \citep[][]{kolarik2020comparing}{}{}, which we were limited to due to GPU memory constraints. The encoder block reduces the spatial dimension while increasing the feature channels progressively, and the decoder block does the inverse operation.  In our U-Net, we used two blocks per stage.

\begin{figure*}
    \centering
    \includegraphics[width=1\textwidth]{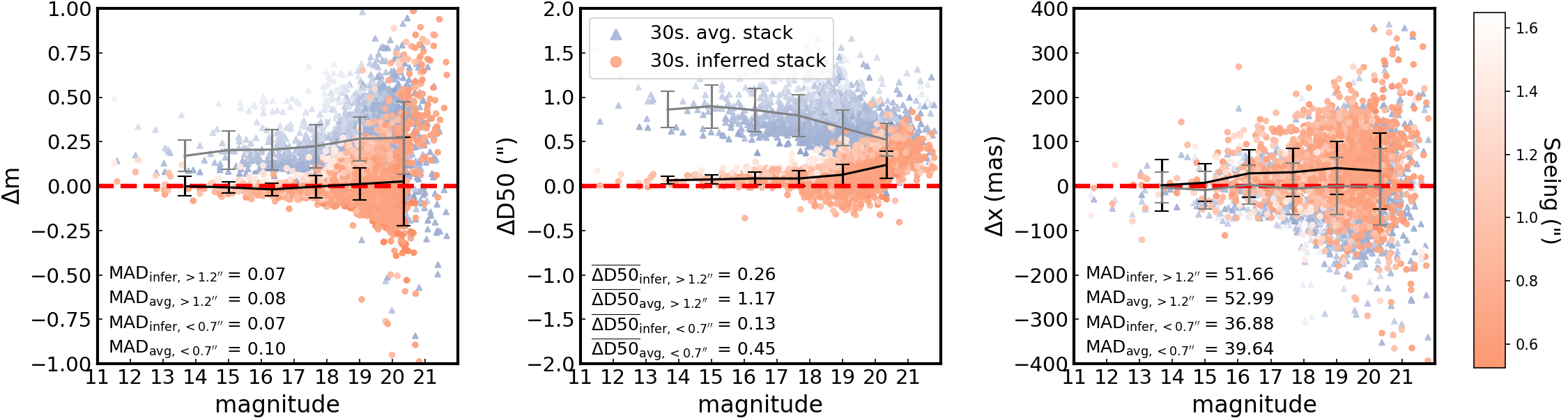}
    \caption{A series of quality assurance tests were made to validate the image reconstructions made by the U-Net. Hundreds of 30-second simulated observations of random stellar fields, with varying seeing conditions, were created and two images were made for each example: a stack made from the U-Net inferred images and a simple averaged stack of the raw frames. \texttt{SExtractor} was run on each frame, along with their corresponding ground truth frame, to collect information about each detected star's (\textit{left panel}) magnitude -- using a 10 pixel diameter aperture -- to test for flux conservation, (\textit{middle panel}) source size -- defined as the diameter of the aperture within which 50\% of the light from a star is contained (D50) -- to test for improvements in image quality, and (\textit{right panel}) centroid coordinates, to test for astrometric stability. Shown here are the residuals of those metrics for the inferred stack (red circles) and simple averaged stack (blue triangles) when compared to the matching stars in the ground truth frames as a function of magnitude, along with their binned means and standard deviations (shown as error bars) -- where the black and grey lines correspond to the inferred and averaged stack values, respectively. Also shown are the computed means for ``bad seeing" and ``good seeing" subsets of the data ($>$\,1.2" and $<$\,0.7", respectively). The ``fainter fatter" effect can be seen for the inferred stacks in the D50 figure, wherein the U-Net tends to smooth the fainter stars (see Section \ref{section:limitations} for a further discussion).
    }
    \label{fig:qualityassurance}
\end{figure*}

The Residual U-Net uses \textit{residual} blocks in the encoder to help alleviate the vanishing gradient problem \citep[][]{he2015deep, borawar2023resnet}{}{} and allow for deeper networks. For an input feature map $X$, each residual block in the architecture performs two convolutional operations $F_1$ and $F_2$, separated by a \texttt{LeakyReLU} activation function. The output of these operations is added to the input feature map, and another \texttt{LeakyReLU} function is applied. This can be mathematically represented as:

\begin{equation}
Y = g_2(F_2(g_1(F_1(X))) + W(X))
\end{equation}
where $F_1$ and $F_2$ denote 2D convolutional operations, $W$ is a transformation operation that adjusts the input feature map's dimensions and/or number of channels to match $F_2(g_1(F_1(X)))$, and $g_1$ and $g_2$ represent the \texttt{LeakyReLU} functions. Each layer consists of two such residual blocks along with down- or upsampling operations. 

Following the principle of \textit{deep supervision} \citep[][]{wang2015training, li2022comprehensive}{}{}, which enables the backpropagation of gradients from the deeper layers of the network to the earlier layers more effectively, our approach compares the output from each decoder block with a correspondingly downsampled version of the ground truth. The loss is calculated for each comparison, and the overall loss is determined by a weighted sum of these individual losses; the weight assigned to each downsampled image is set to be half of that of the preceding image in the sequence. The entire network is shown schematically in Figure \ref{fig:unet}.

\subsubsection{Training configuration}

During training, as a batch of data is loaded, the background is first subtracted by iteratively determining the median and excluding points that deviate more than three times the Mean Absolute Deviation (MAD) from the median. Such an approach effectively minimizes the impact of outliers, such as stars, in the estimation of a statistically robust background level, and ensures the varied background levels of any test data will not negatively influence predictions. 

When working with image reconstruction or denoising, it is often beneficial to transform the input data such that the noise level is approximately constant across the image. This can help algorithms, like those used in U-Net architectures, to perform more consistently. We performed experiments with cross-validation on various scaling transformations. After evaluating the results, in particular the flux conservation, we converged on the Anscombe transformation for both the input and target images. Not only does it help stabilizing the variance, making the noise homoscedastic in the Poisson regime, but it also amplifies signals in dim regions, enhancing the visibility of faint features. The transformation is given by:
\[
f(x) = 2\sqrt{x + \frac{3}{8}}
\]

For its inversion, we adopt the closed-form approximation of the exact unbiased inverse as defined in \cite{makitalo2011closed}:
\[
f^{-1}(x) = \frac{1}{4} x^2 + \frac{1}{4} \sqrt{\frac{3}{2}} x^{-1} - \frac{11}{8} x^{-2} + \frac{5}{8} \sqrt{\frac{3}{2}} x^{-3} - \frac{1}{8}
\]

The application of the Anscombe transform in the context of Deep Supervision, particularly during downsampling of the target frames, presents a methodological concern. Downsampling is typically more effective on linear, untransformed data, yet our approach involves downsampling on Anscombe-transformed data. This raises questions about the potential need for an alternative method, such as downsampling untransformed data and then comparing it to de-transformed outputs from the U-Net, while also considering the implications for the U-Net decoder's handling of dynamic range transformations at each resolution. Alternative strategies will be considered in future iterations of the proposed method; as it stands, the current implementation was deemed sufficient. 

For the optimization process, we employed the Adaptive Moment Estimation (Adam) optimizer, a popular first-order stochastic gradient descent algorithm using estimates of the first and second moments of the gradients \citep{kingma2014adam}. In our implementation, the default values were used such that the initial learning rate was set to 0.001, and beta values of 0.9 and 0.999 were used for the exponential decay rates of the gradient and squared gradient, respectively.

To expedite the training process, we utilized multiple GPUs -- specifically 32GB NVIDIA Tesla V100s -- made available by the Digital Research Alliance of Canada on their cluster. On average, each epoch took approximately 14 minutes to complete (noting that a forward pass of a single example took fractions of a second). The mean-squared error (MSE) loss was used, and a learning rate scheduler was incorporated which adjusted the learning rate when the training MSE loss plateaued.

Since we were particularly concerned with the image reconstructions conserving the flux of stars, \texttt{SExtractor} \citep[][]{bertin1996sextractor}{}{} was run on the inferred and ground truth images in the validation set after every epoch to track the magnitude estimates of detected stars. The final model was chosen such that the mean absolute error between the magnitudes of the inferred and ground truth stars was minimized. We note that astrometric precision was not a criterion during training, but including it in future iterations of the method may help improve overall astrometric performance.


%% file: Sections/Data/data.tex
\section{Data}
\label{section:data}

\subsection{Synthetic data generation \label{section:synthdata}}

Each stellar field of a video sequence in a training dataset was created with one of two methods:

\begin{enumerate}
    \item \textit{Homogeneous field}: The number of stars to be simulated is uniformly sampled from [3, 2000] and placed on a given uniform randomly generated x and y coordinates. An exponential law is used to generate magnitudes for each star: $p(m) \propto 10^{\alpha (m - m_{\rm max})}$, where $m_{\rm max}$ was 21 in this study and $\alpha$ (the slope of the differential source counts, i.e. dlog~N~/~dmag) was 0.4. The inclusion of homogeneous fields helps the model generalize better by learning to handle a diverse set of scenarios with uniform sampling of dense and sparse fields.  

    \item \textit{Realistic field}: To include more realistic priors in our simulation pipeline, we generate the fields according to the distributions of real star clusters of the Milky Way. The catalog from \citet{kharchenko2013global} includes the sky coordinates of thousands of clusters, which we use to query the Gaia DR3 database \citep{prusti2016gaia, vallenari2022gaia} to obtain the RA, Dec, and \texttt{phot\_g\_mean\_mag} values of each star. A random offset in both RA and Dec is uniformly sampled from [$-FOV/2, FOV/2$], where $FOV$ is the field-of-view of the telescope, and applied to get a shifted image, i.e. a more diverse training set. 
\end{enumerate}

The dataset therefore contains both randomly generated and more realistic stellar fields, noting that the inclusion of extended objects (like galaxies) will be included in future iterations of the method (see Section \ref{section:limitations}). 


We generated a training dataset containing 40,000 12-second video sequences; 12 seconds was chosen as a compromise between GPU memory constraints and collecting enough information about the turbulence and faint stars. Each frame is 256x256 pixels and, to match the properties of the C2PU Telescope (see Section \ref{section:c2putelescopedata}), we used a 1-metre diameter telescope aperture, a central wavelength of 650\,nm, a pixel scale of 0.235"/pixel, a readout noise standard deviation of $1e^-$, and a sampling rate of 5.25s$^{-1}$ (i.e. sampled every $\sim$200\,ms) -- noting that while this sampling rate is generally insufficient at capturing the quickly evolving turbulence, it does offer an advantage of having higher signal-to-noise in each frame, and in any case we plan on increasing the sampling rate in future iterations of the method. In each video sequence, the Fried parameter and wind speeds for each layer of the atmosphere are sampled from a normal distribution. Along with each video sequence, we generated the corresponding ground truth frame in which we disabled contributions from the atmosphere and any sources of noise in our simulation pipeline. The datasets were split 90\%/10\% into training/validation sets.

\subsection{C2PU Telescope \label{section:c2putelescopedata}}

The Centre Pédagogique Planète et Univers (C2PU) facility \citep{bendjoya2012c2pu} is located on the Plateau de Caussols at an elevation of 1260 metres, approximately 50 km from Nice in southern France. The site benefits from good seeing in summer time, with a median of 1.06", down to 0.8" at the end of the nights \citep{aristidi2020seeing}.

On the night of May 27, 2022, we collected around 60 seconds of short-exposure (200\,ms) images of the globular cluster M92 in the SDSS r bandpass from the wide-field camera installed at the prime focus of the C2PU Omicron 1.04m telescope. No guiding was used as the tracking accuracy of the telescope allows for unguided exposures up to a few minutes without image degradation.
The wide field camera provides a $37.6'\times 25.2'$ field of view with excellent image quality, through a three-lens Wynne coma corrector, at a resulting F/3.17 focal ratio.
One important feature of this optical setup for our project is its relatively low obstruction (30\% linear) for a wide-field instrument, which preserves more than 80\% of the central peak intensity of the telescopic point spread function, compared to an unobstructed aperture.
The camera at the C2PU Omicron prime focus is a QHY600Pro equipped with a Sony IMX 455M sCMOS sensor. Despite its modest quantum efficiency at redder wavelengths, this generation of sensors has proven to be competitive for quantitative astronomy \citep{betoule2023stardice, alarcon2023scmos}. The sequence images have a size of 1024x1024 pixels and a pixel scale of 0.235".

%% file: Sections/Results/results.tex
\section{Experimental evaluation}
\label{section:results}

\begin{figure*}
    \centering
    \includegraphics[width=0.9\textwidth]{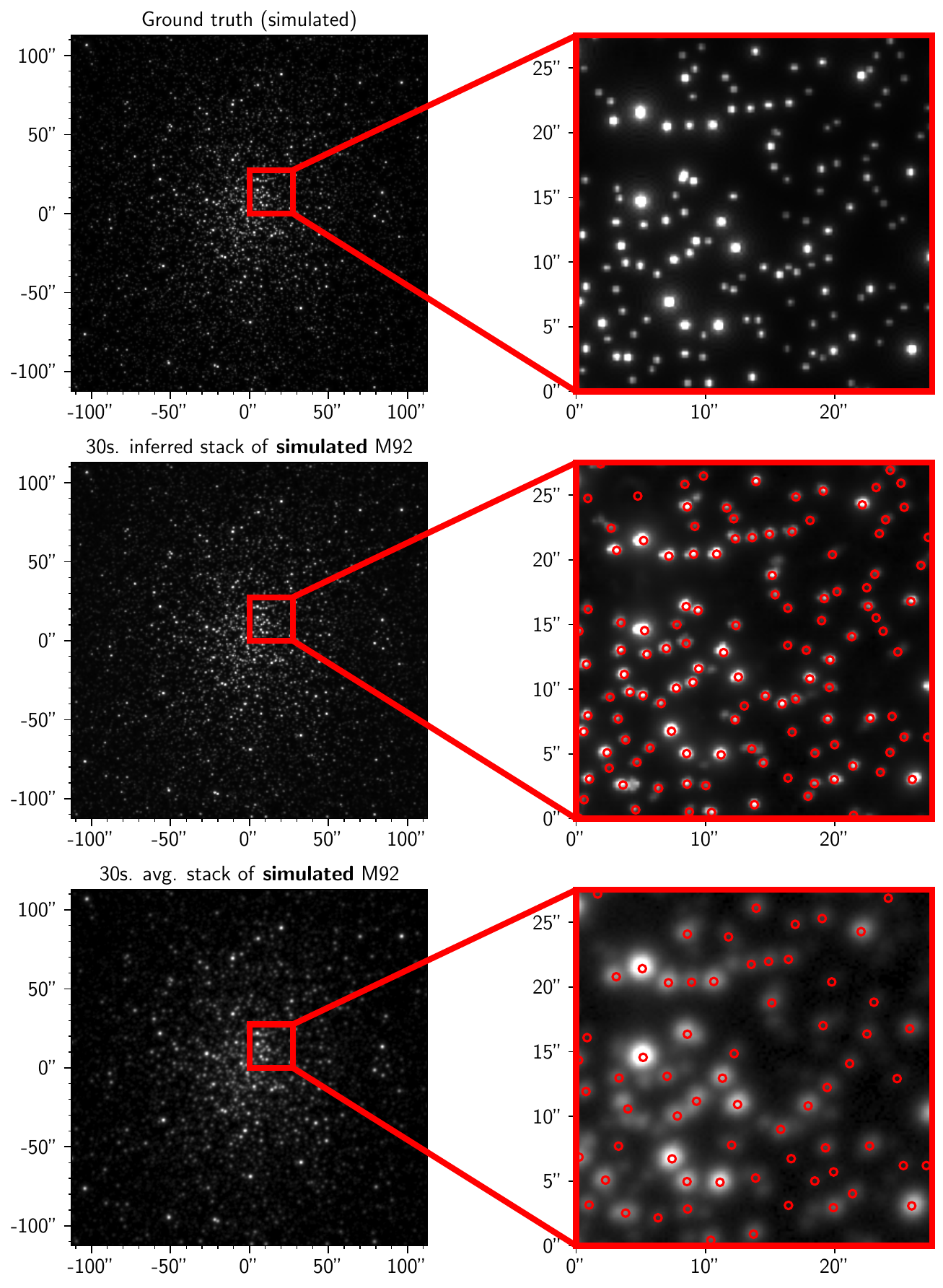}
    \caption{Our simulation pipeline was used to create a 30 second video stream -- 160 frames in total @ 5.25 frames/sec -- of the globular cluster M92 using Gaia positions and G magnitudes and a total seeing of 1.36". Shown here is a comparison of the simulated ground truth, a stack made from inferred images, and a simple averaged stack of the frames. The red circles indicate stars that were detected by \texttt{SExtractor}, using conservative detection settings.}
    \label{fig:M92sim_zoomed}
\end{figure*}

\begin{figure}
    \centering
    \includegraphics[width=0.5\textwidth]{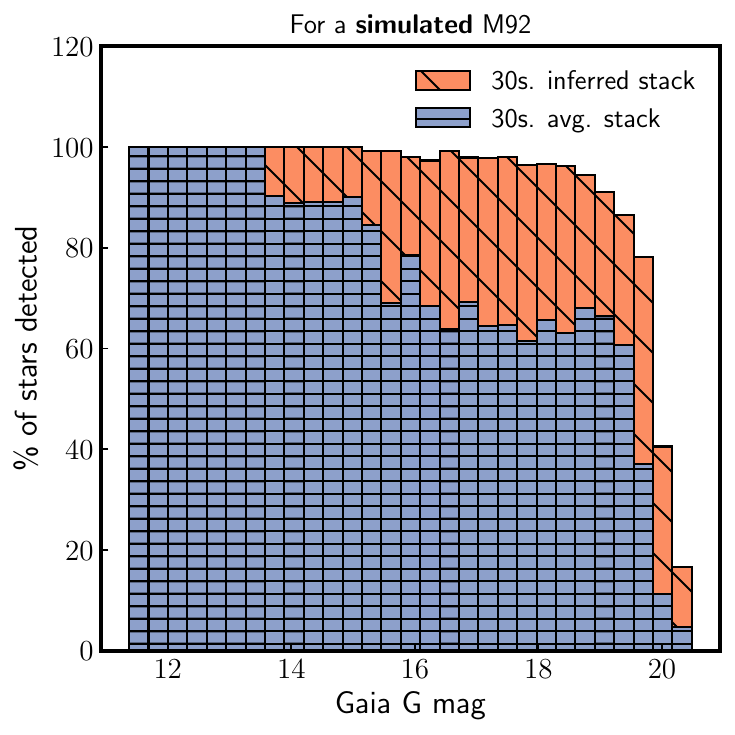}
    \caption{\texttt{SExtractor} was run on the ground truth frame, inferred stack, and simple averaged stack of the simulated M92 shown in Figure \ref{fig:M92sim_zoomed}. The stars that were identified in the inferred stack and averaged stack were matched to the stars in the ground truth frame (by ensuring their measured positions differed by less than 2 pixels), and shown here is the completeness of the detected stars as a function of Gaia G magnitude. Note that the false positive rate for both frames was $\sim$2\%, i.e. of all the stars \texttt{SExtractor} identified in the inferred and averaged frames, 2\% were not matched to those in the ground truth frame.}
    \label{fig:M92sim_percentstarsrecov}
\end{figure}

Visually, the proposed method does an excellent job at taking in a short sequence of turbulent images and producing a clear, sharp, noise- and turbulence-mitigated image. For example, Figure \ref{fig:examples_inferred_goodbadseeing} shows the massive improvement -- compared to temporally averaging the sequence -- in image quality of frames inferred from sequences with good (0.7") and bad (1.4") seeing conditions. Some stars which are barely visible (or not visible at all) in the averaged frame appear quite clearly in the inferred frame.

Validating the method \textit{quantitatively} in addition to qualitatively, however, required analyzing the performance metrics of a large number of simulated observations, as detailed in the following section. 

\begin{figure*}
    \centering
    \includegraphics[width=1\textwidth]{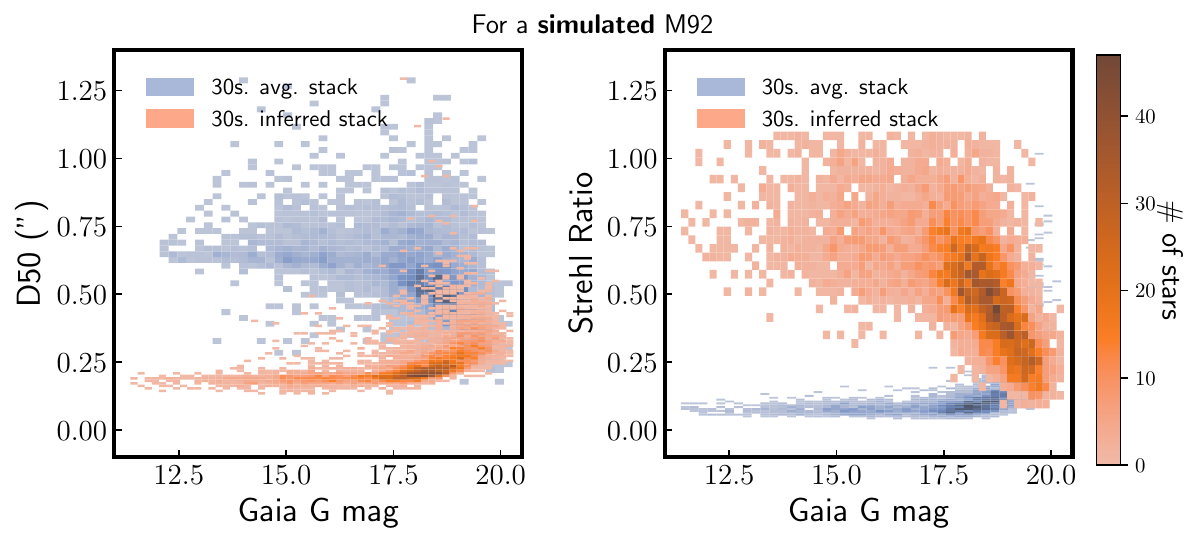}
    \caption{\texttt{SExtractor} was run on the inferred stack and simple averaged stack of the simulated M92 shown in Figure \ref{fig:M92sim_zoomed}. Shown here are the distributions of the angular diameter inside of which 50\% of the flux is contained (D50, left plot) and Strehl ratio estimates (right plot) of the detected stars as a function of Gaia G magnitude.}
    \label{fig:M92sim_d50_strehl}
\end{figure*}
 
\subsection{Quality assurance}
\label{section:qualityassurance}

To quantitatively assess the performance of the Residual U-Net, a series of quality assurance tests were implemented. These tests were designed to evaluate the improvements in image quality, flux conservation, and astrometric stability in comparison to simply stacking the images. 

A 30-second video sequence of a stellar field was simulated following the \textit{random field} process described in Section \ref{section:synthdata}, including the random sampling of Fried parameters and wind speeds for each atmospheric layer. We iteratively step along the sequence one frame at a time and partition a 12-second subset to obtain an inferred U-Net frame until a full sequence of inferred frames was collected. The inferred sequence was stacked by taking the mean along the temporal dimension, resulting in a single frame representing a 30-second inferred observation. The same 30-second sequence was used to obtain a single averaged frame. We additionally saved the ground truth frame (simulated with no turbulence or noise) for each sequence. The entire procedure was repeated 500 times, resulting in a ``quality assurance" dataset of 500 inferred frames, 500 averaged frames, and 500 ground truth frames. 

\texttt{SExtractor} was then used on all of the frames in the dataset to identify the stars\footnote{For the sake of consistency across the vastly different regimes of source density and image quality, in this work we used fixed, conservative source extraction settings for \texttt{SExtractor} except where otherwise noted.} and calculate for each one the magnitude (by using an aperture diameter of 10 pixels to measure the total flux), the angular diameter (in terms of the D50, which measures the diameter at which 50\% of the light is contained within an aperture), and the x-position, for the purpose of showing flux conservation, improved image quality, and astrometric stability, respectively. Figure \ref{fig:qualityassurance} shows the residuals of these estimates, as a function of magnitude, on the inferred and averaged frames compared to the ground truth frames.

The magnitude estimates of stars from the inferred frames can be seen to have overall a smaller dispersion throughout the magnitude range than stars in the averaged stack (which is evidenced by the median absolute deviation (MAD) being 0.01"\,-\,0.03" lower across the seeing range). It is also the case that the magnitude residuals are closer to ground truth for the inferred frames, but in practice this could easily be resolved by using an adaptive aperture size on the averaged frames; we emphasize the importance of the magnitude residual plot is showing that flux is conserved in the inferred frames with lower dispersion than for the averaged frames. The lowered dispersion is also shown in the astrometric stability plot, where the MAD is 1\,-\,3\,mas less for the inferred frames than the averaged frames, a very modest reduction but importantly there is no apparent degradation in astrometric stability. The strength of our proposed method is showcased in the angular diameter plot, where the average D50 residual is a factor of 3\,-\,4x lower in the inferred frames than the averaged frames, and for bright stars can even be a factor of about 8x less.



In all cases, the performance on the inferred frames gets worse towards fainter magnitudes: at a magnitude of 16, the standard deviations for magnitude residuals and D50 residuals are 0.01 and 0.07", respectively, whereas at a magnitude of 19, the residuals increase to 0.08 and 0.1". It appears the U-Net tends to ``smooth" fainter stars, leading to a "fainter fatter" effect (as opposed to the more common "brighter fatter" effect for CCDs), further discussed in Section \ref{section:limitations}.


\subsection{Test case: M92}
\label{section:testcaseM92}

After ensuring that the trained model performed well in conserving flux, increasing image quality, and stabilizing the astrometry, it was time to task it with a more realistic scenario: inferring the turbulence-free image of the globular cluster M92. With the high stellar densities in the core of the cluster, this would be a true test of our method's ability to disentangle the light from nearby stars and improve the spatial resolution. Testing was performed on both simulated data and real data obtained with the C2PU telescope. 

\subsubsection{Simulated data}
\label{section:simM92results}

We began the test case of M92 with a simulated observation of it (30 seconds total @ 5.25 frames/second for a total of 160 frames) to obtain a baseline to compare to. A wide field of view ($\sim$4\,arcmin, 1024$\times$1024 pixels) was used to showcase the ability of our method to perform on larger images than it was trained on. As described in Section \ref{section:synthdata}, Gaia coordinates and magnitudes were used for each star. We set a maximum magnitude of $G=21$, consistent with the model's training parameters, leading to a total of $\sim$12,000 stars being simulated. A relatively poor seeing of 1.36" was chosen to test the method's proficiency under sub-optimal conditions. 

Since the model was trained on frames of size 256$\times$256 pixels, the full 1024x1024 frames of M92 could not be used as input. Instead, each frame was split into tiles of size 256$\times$256 with 50\% overlap, resulting in the full 1024$\times$1024$\times$160 video sequence being split into 49 overlapping sequences of size 256$\times$256$\times$160. As described in Section \ref{section:qualityassurance}, the U-Net was used on each of these sequences to produce a series of inferred images along the temporal axis which were then averaged together to produce a single 256$\times$256 inferred image, thereby reducing the 49 sequences of size 256$\times$256$\times$160 into overlapping inferred tiles of size 256$\times$256. In the final reconstruction, the overlapping regions of these tiles were blended smoothly: masks were generated for each tile, consisting of a central region with full contribution (value 1), and edge regions with a linear gradient from 0 to 1, reflecting the degree of overlap. The final image \( B \) at each pixel was reconstructed by calculating the normalized weighted sum of the $n$ overlapping tiles, expressed as \( B(x, y) = \frac{\sum_{i=1}^{n} T_i(x, y) \cdot M_i(x, y)}{\sum_{i=1}^{n} M_i(x, y)} \), where \( T_i(x, y) \) and \( M_i(x, y) \) are the value and mask at position \( (x, y) \) in the \( i^{th} \) tile, respectively. This ensured a seamless integration of the tiles, with the centre of each tile retaining its original value and the overlapping edges merging smoothly, resulting in the final 1024$\times$1024 inferred image.

Figure \ref{fig:M92sim_zoomed} shows the ground truth frame, the inferred frame, and the temporally averaged frame, along with zoomed in images of the central region of M92 which show, with red circles, stars identified with \texttt{SExtractor}. It is visually clear that the inferred frame is far sharper and had significantly more stars identified than the averaged frame, so a more quantitative analysis was conducted to confirm this. 

Figure \ref{fig:M92sim_percentstarsrecov} shows the relative number of stars identified by \texttt{SExtractor} in both the inferred and averaged frames in bins of magnitude, where true positives were defined as being no more than 2 pixels away from the corresponding star in the ground truth frame. We note that the detection threshold for \texttt{SExtractor} was optimized on each frame individually to maximize both the precision and recovery rate of its star identification. For the averaged frame, the precision was 97.1\% (i.e. 2.9\% false positive rate), and 63.1\% of stars were successfully recovered. In contrast, for the inferred frame, the detection precision was 97.8\% (i.e. 2.2\% false positive rate), and 86.2\% of stars were successfully recovered, verifying that substantially more stars were identified in the inferred frame. 



To test for the improvements in image quality, the D50 measurements from \texttt{SExtractor} were used and, because it is a more common diagnostic metric in AO analyses, the Strehl ratio, S, was calculated for each star according to \cite{roberts2004really}:

\begin{equation*}
S = \frac{I(x=0)}{\sum{I}} \frac{\sum{P}}{P(x=0)},
\end{equation*}
where $x$ is the position vector, $I(x=0)$ is the maximum of the measured PSF, $P(x=0)$ is the maximum of the diffraction limited PSF, and $\sum{I}$ and $\sum{P}$ are computed over squares of size 10$\times$10 pixels (centred on the coordinates extracted by \texttt{SExtractor}) and used to normalize the ground truth PSF to have the same total intensity as the observed PSF. This Strehl ratio estimator tends to be noisy due to the statistical noise and crowding, but we are more interested in \textit{relative} improvements between the average and inferred stacks. Figure \ref{fig:M92sim_d50_strehl} shows how these values change over the magnitude range. Below a Gaia G magnitude of 17.5, the inferred stars have a 3x average reduction in D50 and an average of $\sim$6x improvement in Strehl ratio. Beyond a magnitude of 17.5, the performance on inferred stars drops yet still maintains a 1.5-3x improvement on D50 measurements, and 2-4x improvement on Strehl ratios; again we find that the model struggles with fainter stars, tending to smooth their PSFs.

\subsubsection{C2PU telescope data}

\begin{figure*}
    \centering
    \includegraphics[width=0.9\textwidth]{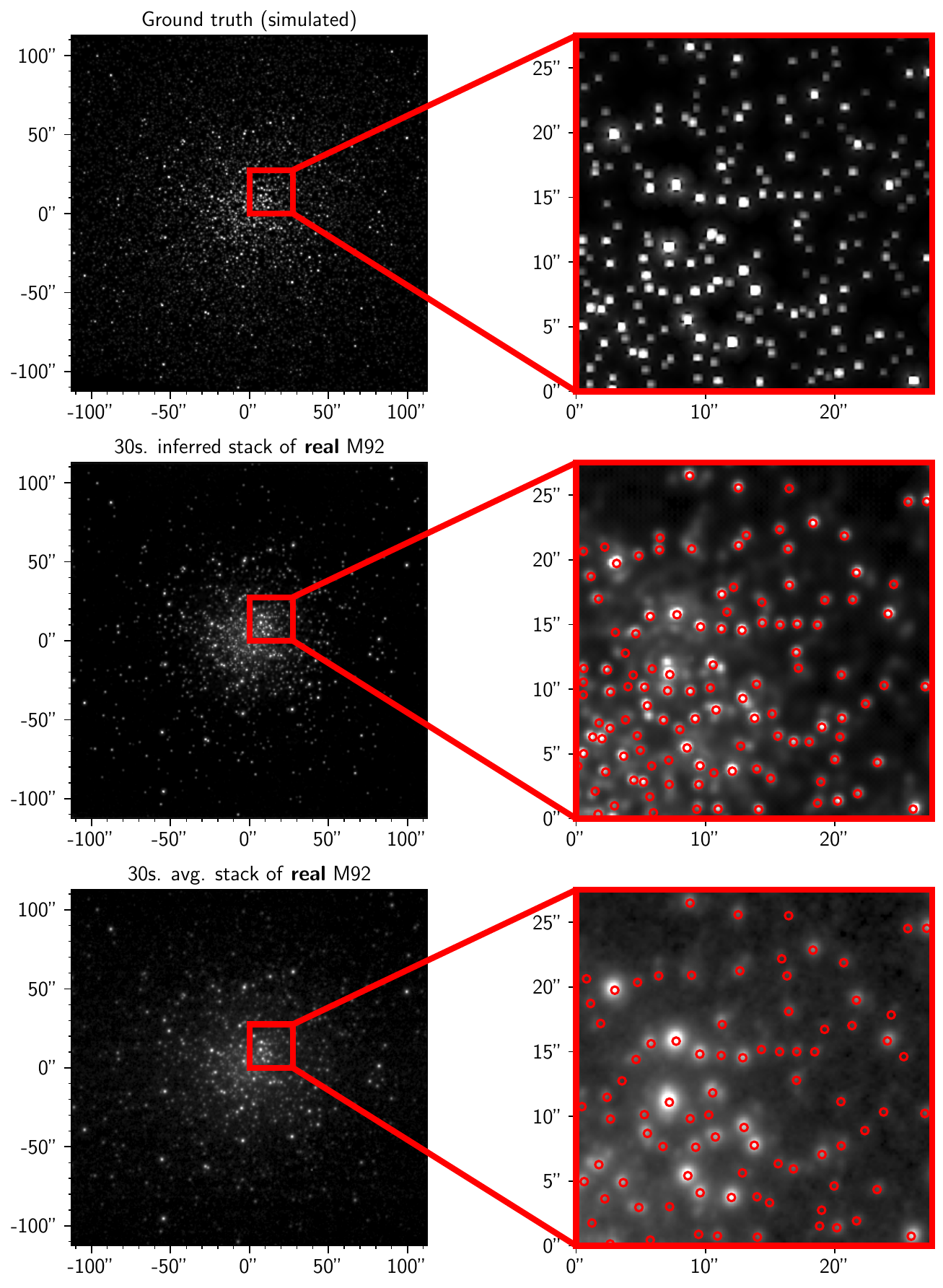}
    \caption{The C2PU telescope was used to obtain a 30 second video stream -- 160 frames in total @ 5.25 frames/sec -- of the globular cluster M92. Shown here is a comparison of the simulated ground truth using Gaia positions and magnitudes, a stack made from inferred images, and a simple averaged stack of the frames. The red circles indicate stars that were detected by \texttt{SExtractor} using conservative detection settings. Note that some relatively bright stars are missing from the Gaia catalog in these crowded areas.}
    \label{fig:M92_zoomed}
\end{figure*}

All of the steps in \ref{section:simM92results} were repeated for the real M92 data collected by the C2PU telescope. Figure \ref{fig:M92_zoomed} shows a comparison between the ground truth (simulated with Gaia coordinates and magnitudes), inferred, and temporally averaged images of a 30 second observation of M92 containing 160 total frames of size 1024x1024. The zoomed in areas of the central regions of M92 show that the brighter stars in the inferred image appear substantially sharper, though there are clearly some spurious effects in the fainter parts; in the video stream of M92, there were indications that an intermittently very turbulent ground layer was causing all the stars in the field to move in lockstep by up to a couple arcseconds, an effect which is not currently accounted for in the simulations. This and other limitations, as discussed in Section \ref{section:limitations}, contributed to a decreased visual performance on real data.  

\begin{figure}
    \centering
    \includegraphics[width=0.5\textwidth]{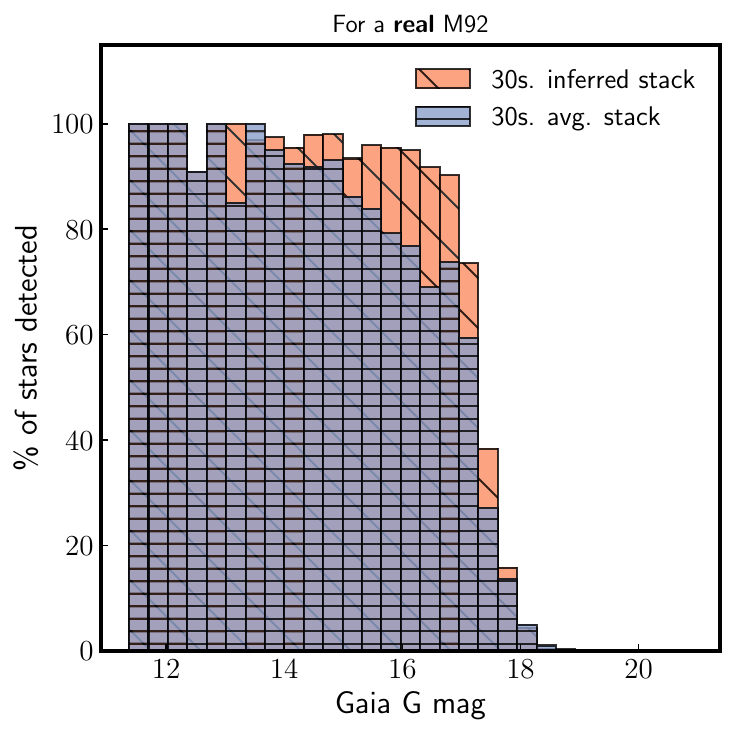}
    \caption{\texttt{SExtractor} was run on the ground truth frame, inferred stack, and simple averaged stack of the real M92 shown in Figure \ref{fig:M92_zoomed}. The stars that were identified in the inferred stack and averaged stack were matched to the stars in the ground truth frame, and shown here is the completeness of the detected stars as a function of Gaia G magnitude. Note that most of the stars in M92 are fainter than G\,$\sim$\,18, so the majority of stars not detected had G\,$>$\,18.}
    \label{fig:M92_percentstarsrecov}
\end{figure}

\begin{figure}
    \centering
    \includegraphics[width=0.5\textwidth]{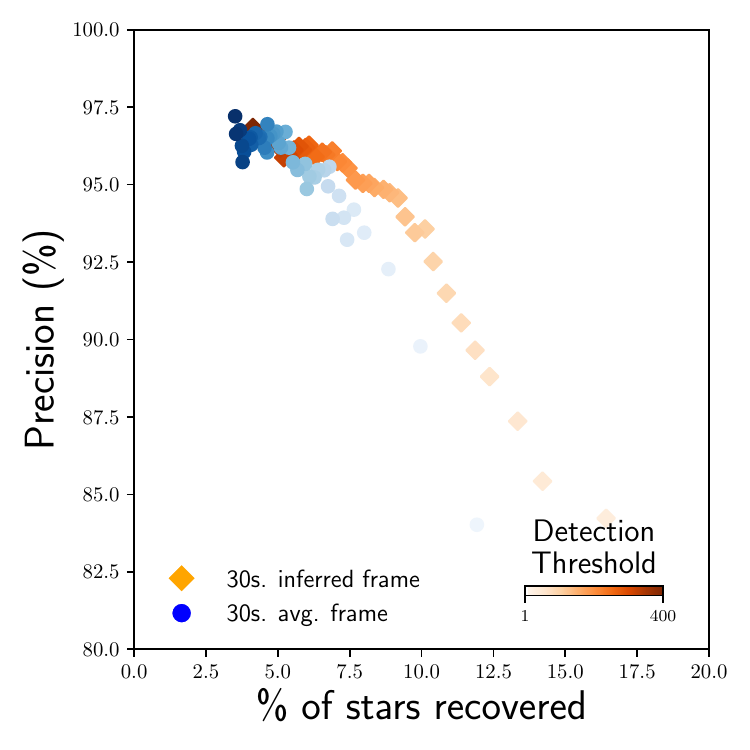}
    \caption{The chosen detection threshold of \texttt{SExtractor} affects the precision of detected sources and total percentage of recovered stars. Shown here are the results of changing the detection threshold on a 30 second averaged frame and a 30 second inferred frame of M92.}
    \label{fig:M92_precisionrecall}
\end{figure}

\begin{figure}
    \centering
    \includegraphics[width=0.5\textwidth]{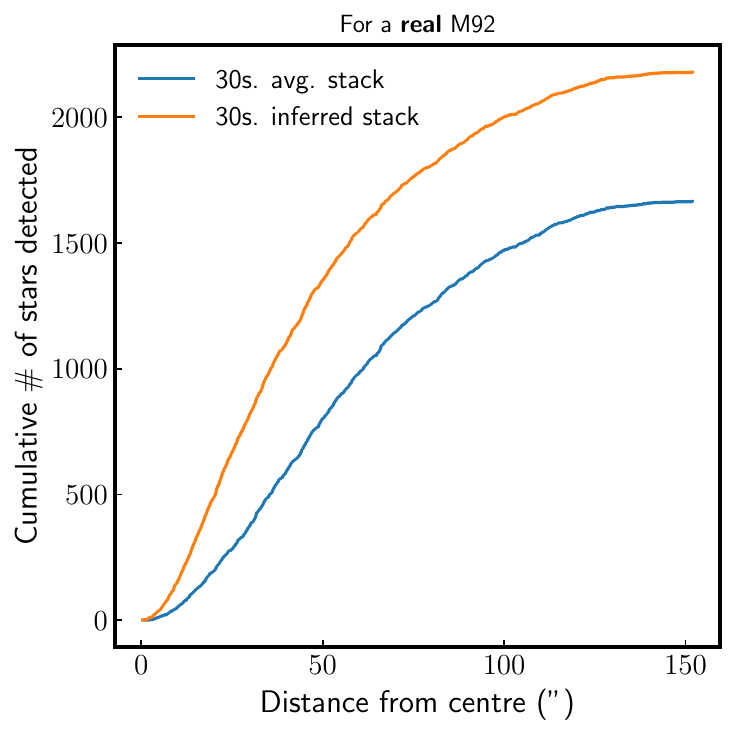}
    \caption{\texttt{SExtractor} was run on the inferred and simple averaged stack of the real M92 shown in Figure \ref{fig:M92_zoomed}. Shown here is the cumulative number of stars found as a function of distance from the centre of the cluster for each stack, highlighting the ability of the proposed method to work in the crowded central regions of a stellar cluster.}
    \label{fig:M92_cumulativestars}
\end{figure}

As for the number of stars correctly identified, Figure \ref{fig:M92_percentstarsrecov} shows that slightly more stars -- throughout the entire magnitude range -- were found in the inferred frame than the averaged frame ($\sim$9.6\% vs. $\sim$8.7\% of the total stars in the ground truth image), however the total number of stars recovered in either frame, as well as the precision of the detections, depends on the detection threshold used in \texttt{SExtractor}. Figure \ref{fig:M92_precisionrecall} shows the relationship between the precision and total number of stars correctly identified as a function of detection threshold, where it can be seen that better performance (a maximum of a few percent more stars recovered for a given precision) is achieved for the inferred frame. We also investigated \textit{where} the stars were being more readily identified in both images. Figure \ref{fig:M92_cumulativestars} shows that the inferred image works particularly well at de-blending the crowded central regions of M92, recovering about 3x more stars within 25\,arcsec from the centre than the averaged image.    

\begin{figure*}
    \centering
    \includegraphics[width=1\textwidth]{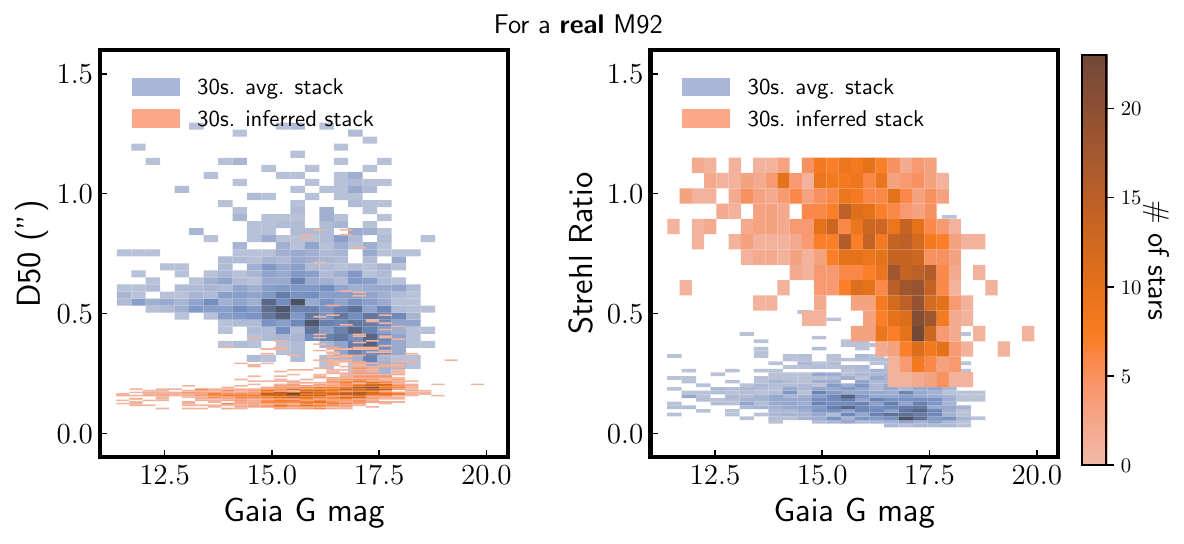}
    \caption{\texttt{SExtractor} was run on the inferred stack and simple averaged stack of M92 shown in Figure \ref{fig:M92sim_zoomed}. Shown here are the distributions of D50 (left plot) and Strehl ratio (right plot) estimates of the detected stars as a function of Gaia G magnitude.}
    \label{fig:M92_d50_strehl}
\end{figure*}

Finally, Figure \ref{fig:M92_d50_strehl} shows quantitatively how the D50 and Strehl ratio measurements on the inferred and averaged images change across the magnitude range. Again we find that, for the inferred image, the performance on brighter stars (G\,$<$\,17) is enhanced, with an average of 2.5x reduction in D50 measurements and 5x increase in Strehl ratio measurements. For fainter stars, the performance drops in a similar manner as in the case of a simulated M92, leading to only slightly decreased average PSF width than the brighter stars. 


The results from the analysis on the synthetic and observed M92 are summarized in Table \ref{tab:m92_comparison}, including the results from using a smaller temporal input context for training and testing the U-Net (described in Section \ref{section:appendix_smaller_context}). Additionally, results from training on single temporally averaged frames are discussed in Section \ref{section:avgframeinference}.

\begin{table*}
\centering
\caption{Comparison of the precision, percentage of recovered stars, average angular diameter ($\overline{\textrm{D50}}$), and average Strehl ratio ($\overline{\textrm{S}}$) for the various image processing methods used for simulated and real M92 observations. We note that the chosen detection threshold in \texttt{SExtractor} can affect the precision and recall substantially (see Figure \ref{fig:M92_precisionrecall}). In the case of real M92 observations the detection threshold for the averaged stack could be chosen such that the precision was fairly close to 100\% but with a very small total number of stars detected, whereas the inferred stacks had a definite maximum precision. For a more direct comparison, the thresholds for the averaged stack were chosen such that their resulting precision was approximately equal to the maximum precision for the inferred stack. }
\label{tab:m92_comparison}
\begin{tabular}{|l|c|c|c|c|c|c|c|c|}
\hline
\textbf{Method} & \multicolumn{4}{c|}{\textbf{Simulated M92 Results}} & \multicolumn{4}{c|}{\textbf{Real M92 Results}} \\ \hline
 & \textbf{Precision (\%)} & \begin{tabular}[c]{@{}l@{}}\textbf{Recovered}\\ \textbf{Stars (\%)} \end{tabular} & \textbf{$\overline{\textrm{D50}}$ (")} & \textbf{$\overline{\textrm{S}}$} & \textbf{Precision (\%)} & \begin{tabular}[c]{@{}l@{}}\textbf{Recovered}\\ \textbf{Stars (\%)} \end{tabular} & \textbf{$\overline{\textrm{D50}}$ (")} & \textbf{$\overline{\textrm{S}}$} \\ \hline
30s. Avg. Stack & 97.1 & 63.1 & 0.63 & 0.13 & 97.1 & 8.73 & 0.50 & 0.20 \\ \hline
\begin{tabular}[c]{@{}l@{}}30s. U-Net Inferred Stack\\ (6s. input context)\end{tabular} & 96.4 & 85.8 & 0.32 & 0.51 & 97.2 & 10.04 & 0.24 & 0.52 \\ \hline
\begin{tabular}[c]{@{}l@{}}30s. U-Net Inferred Stack\\ (12s. input context)\end{tabular} & 97.8 & 86.2 & 0.23 & 0.60 & 97.2 & 9.57 & 0.19 & 0.74 \\ \hline
\end{tabular}
\end{table*}

%% file: Sections/Discussion/discussion.tex
\section{Discussion}
\label{section:discussion}

The proposed method introduces a novel approach to handling the challenges posed by atmospheric turbulence in astronomical imaging, especially when dealing with wide fields. By segmenting the observed field into smaller overlapping tiles, producing an inferred frame for each tile, and merging the inferred tiles, the method allows for an effective mitigation of atmospheric turbulence effects in arbitrarily large images without compromising the quality of turbulence correction.

An important feature of the DanceCam approach is its independence from the traditional reliance on guide stars, instead using the information from the entire field to understand the turbulence properties. While AO systems are tethered to either natural or artificial guide stars for measuring and correcting atmospheric distortions, our method's flexibility broadens its applicability, offering the potential to correct turbulence even in regions where suitable guide stars might be elusive. Moreover, the ability to analyze distortions in captured images paves the way for reverse-engineering the atmospheric layers and their respective turbulence properties. Such insights can be beneficial for our understanding of atmospheric conditions and refining observation strategies.

Modern technological advancements further strengthen the efficacy of the method. With the requirement of short exposures to capture the near real-time turbulence, most CCDs would produce images with too much readout noise, effectively drowning out the faint targets. The integration of sCMOS cameras, on the other hand -- renowned for their rapid readout speeds, reduced noise, and enhanced quantum efficiency \citep{guidash2016reduction, zhang2020design, zhu2022design} -- complements the turbulence mitigation technique. Modern GPUs, additionally, allow for an impressive processing speed; the time required to produce a single inferred frame from a fully trained model is mere fractions of a second, allowing for the correction of a video stream comprising hundreds of short-exposure wide-field images in near real-time.

The proposed method is also appealing due to its minimalistic hardware requirements. Eschewing the intricate setups of traditional AO systems, which involve expensive deformable mirrors and wavefront sensors, as well as significant overheads, our approach is predominantly software-centric. This orientation not only simplifies its implementation but also offers significant cost savings, making it an attractive solution for budget-conscious observatories and researchers, and even amateur astronomers.

\subsection{Limitations and Known Issues}
\label{section:limitations}

Acknowledging the current limitations and exploring potential avenues for improvement of the proposed method is essential for its efficacy and ensuring its applicability in future astronomical observations:

\begin{itemize}
    \item \textbf{Intermittent Turbulence:} One of the primary constraints is the absence of intermittent turbulence in the current implementation. Intermittent turbulence can introduce sporadic and unpredictable distortions, which the method might not handle effectively in its present form. Interestingly, it could even enhance the method's effectiveness by occasionally providing ``lucky frames" in a sequence. A straightforward solution could involve incorporating dynamic changes to $r_0$ for each atmospheric layer in the simulation pipeline.

    \item \textbf{Motion Blurring:} Another limitation is the lack of motion blurring implementation. In real-world scenarios, the shutter collects light over a duration (e.g., 0.2 seconds), whereas the simulation pipeline currently takes instantaneous snapshots at regular intervals. Addressing this could involve splitting each frame into N sub-frames, simulating turbulence at every sub-timestep, and averaging the sub-frames together. This approach would effectively simulate an open shutter, offering a more realistic representation. Alternatively, using shorter exposures could also address this limitation.

    \item \textbf{Monochromatic Light:} The method's current reliance on simulating monochromatic light simplifies the computational process but overlooks chromatic aberrations introduced by atmospheric turbulence. Enhancing the simulation pipeline to cover a broader band-pass of light could improve the method's accuracy.

    \item \textbf{Readout noise:} In the current scheme, we assume that readout noise follows a Gaussian distribution. However in sCMOS sensors each pixel has its own amplifier circuit, and therefore its own noise characteristics; some pixels are markedly noisier than others, and the overall histogram of a real uniformly illuminated sCMOS exposure is significantly non-Gaussian.

    \item \textbf{Rolling shutter:} Today's astronomical sCMOS sensors are usually operating in rolling shutter mode: lines of pixels are read and reset in sequence. The efficiency of photon collection is thereby optimized, but this has the inconvenience that two successive sensor lines are recorded at slightly different times. The maximum time difference amounts to only a few milliseconds inside the current $256\times 256$ analysis window. However this delay would reach $\sim$100ms if distant parts of the full sensor were to be analyzed jointly, in which case it could not be neglected anymore.

    \item \textbf{GPU Constraints:} Temporal context is currently limited due to GPU memory constraints. Optimizing GPU usage, such as parallelizing the pipeline to distribute computational load across multiple GPUs, could address this limitation. Additionally, as GPUs continue to technologically advance, the method's capabilities will naturally expand.

    \item \textbf{Only simulating stars:} A major limitation of the method is that it is currently tailored to simulate stars only, excluding other celestial objects like nebulae, cirri, galaxies or planets. Integrating diverse light profiles and spatial structures associated with a variety of astronomical objects would enhance its scope and versatility.

    \item \textbf{Transient Objects:} The current system does not handle transient objects that appear temporarily and unpredictably, e.g. satellites or other moving objects, or objects that change in brightness. These types of objects could be included in the simulation pipeline. Alternatively, or additionally, owing to the high cadence video streams, frames containing transients of interest could be detected and saved for later processing using automated methods \citep[e.g., ][]{cabrera2017deep, gieseke2017convolutional}{}{}. Deleterious transients (specifically telecommunications satellites) could be removed before processing. This is indeed a major benefit of the video stream approach to observations, since with long exposures the best you can do is mask satellite streaks \citep[e.g., ][]{paillassa2019maximask}. See \citet{Beskin_Biryukov_Gutaev_Karpov_Oganesyan_Valyavin_Valeev_Vlasyuk_Lyapsina_Sasyuk_2023} for a more in-depth look at the benefit of wide-field, high cadence imaging for transient detection and analysis. 

    \item \textbf{High-Volume Data Streams:} Given our focus on wide fields and short exposures, the method generates a significant volume of data, which can be challenging to process and store efficiently. To manage these massive data streams, it will likely be crucial to process and analyze the collected data in quasi-real-time using one or several GPUs, helping to filter and prioritize data for storage and detailed examination.

    \item \textbf{Circular Stationarity of Phase Screens:} In the current scheme, we initialize a static phase screen and ``roll" the values along for each time step, where the values at the edge roll over to the opposite edge. This approach, while efficient, isn't a problem when using short temporal contexts for training. However, it could become increasingly problematic when using longer video streams, as it introduces artificial periodicity and may not accurately represent the evolving nature of atmospheric turbulence. To address this, a more dynamic model of phase screens could be developed, where new turbulence patterns are continuously generated rather than recycled. This would better mimic the natural, non-repetitive behaviour of atmospheric turbulence over extended periods. Implementing algorithms that can generate realistic, time-evolving turbulence patterns without significantly increasing computational load would be key.
\end{itemize}

In addition to these limitations, there are other effects not accounted for in the simulations, including 16-bit quantization, dark currents, quantum efficiency, malfunctioning pixels, and filter variations. These effects can all be added in future iterations of the method, either in simulations or by using a semi-supervised approach to jointly train on real observations.

Two notable issues that have emerged are the ``fainter fatter" effect, where fainter stars are excessively smoothed, and the ``hallucination" effect, where the U-Net erroneously interprets noise as non-existent stars (as evidenced by imperfect precision of detected sources). These phenomena have multiple potential causes:

\begin{itemize}
    \item Limited Temporal Context: Atmospheric turbulence disperses starlight, reducing the signal-to-noise ratio per pixel, sometimes burying the signal entirely. With limited information, the U-Net's tendency to misinterpret noise (either as part of a faint star's profile or as a new star) increases, especially for faint stars with inherently weaker signals.
    
    \item U-Net's Inherent Design:  The U-Net's output, when trained with a square-error cost, converges to the mean of the posterior distribution. Indeed, the output is a probability-weighted average of  inferred images compatible with the data, where the solution's likelihood is much more ``peaky" for bright sources than for faint sources. 

    \item Difficulty Tracking Faint, Isolated Stars: The U-Net might struggle to consistently ``track" extremely faint, isolated stars.  Extending the temporal context may offer limited improvement for this specific issue.
\end{itemize}

These observed limitations highlight the complex interplay between atmospheric conditions, U-Net architecture, and the challenges inherent in detecting faint objects.



Addressing these limitations could involve expanding the temporal context of the observations, as shown in Table \ref{tab:m92_comparison} and discussed in Section \ref{section:appendix_smaller_context}, where it is confirmed that an increase in temporal context correlates with improvements in precision, Strehl ratio, and D50 measurements. This suggests that a broader temporal scope allows for a more accurate signal reconstruction, particularly for fainter stars.

Future enhancements to the proposed method could benefit from exploring diverse approaches to uncertainty quantification. For example, the method outlined by \citet[][]{angelopoulos2021gentle} offers a framework for creating statistically rigorous, distribution-free uncertainty sets for any pre-trained model's predictions, without relying on distributional or model assumptions. This technique ensures that the generated sets contain the ground truth with a specified probability. In contrast, probabilistic or generative  models, such as Bayesian neural networks, generative adversarial networks (GANs), or diffusion-based models, provide an alternative means of assessing uncertainty. These approaches could generate probabilistic distributions for each pixel, offering a detailed view of the possible inferences and their associated uncertainties. These methods would allow for a more granular understanding of uncertainty, presenting the variability of reconstructions in a quantitatively rich manner, thus offering a distinct advantage in scenarios where understanding the distribution of inferences is crucial.


Another important limitation of our current method is its inability to significantly improve astrometric precision despite sharper images. This shortcoming likely stems from the residual jitter caused by atmospheric turbulence, as the U-Net, though capable of tracking the turbulent PSF across exposures, cannot determine the true position of stars. Averaging the centroids of these wandering PSFs over the 32 or 64 frames is not significantly more beneficial than averaging the light distribution, leaving a gap in astrometric accuracy. The addition of priors coming, for example, from the Gaia catalog could help to remove this limitation.

While the method has its limitations, many of these challenges offer exciting avenues for future research and development. Addressing these constraints can evolve the method into a more robust and comprehensive solution for astronomical imaging in the presence of atmospheric turbulence. Indeed, we would like to emphasize that the work presented in this paper represents only the first iteration of the proposed method. There are numerous optimizations which can be implemented to enhance the performance by, e.g., incorporating more realism into the simulations, broadening the temporal and spatial contexts over which the U-Net makes its inferences, using shorter duration images to capture more information about the turbulence, and implementing more advanced machine learning methodologies and architectures.

%% file: Sections/Conclusions/conclusions.tex
\section{Conclusions}
\label{section:conclusions}

In this study, we introduced a novel machine learning-based approach to counteract the challenges posed by atmospheric turbulence in astronomical imaging. By utilizing a U-Net architecture, we have demonstrated the potential to significantly enhance the sharpness of astronomical images. Our method, trained on simulated observations, is adept at inferring a turbulence- and noise-free image from a sequence of short-exposure observations of a stellar field, effectively associating speckles with their source star and disentangling light from proximate sources, while conserving flux.

Visually, the method showcased an enhancement in image clarity, especially under sub-optimal seeing conditions. Quantitatively, our results have been compelling: when tested on the simulated M92 dataset, the inferred frames exhibited an average reduction in D50 measurements by a factor of 3 for stars brighter than a Gaia G magnitude of 17.5, and an average 6x improvement in the Strehl ratio. This performance, however, tapered for fainter stars, indicating areas for further refinement. Furthermore, our quantitative analysis using \texttt{SExtractor} revealed that, when using the inferred frame, up to 36\% more stars were identified relative to the averaged frame, with a precision rate of $\sim$98\%. This is a testament to the model's ability to enhance image quality and resolution, even in densely populated stellar fields.

In real-world tests, using a 30\,sec. video sequence of the globular cluster M92 as a case study, our method demonstrated its ability to de-blend crowded regions. Specifically, the inferred image recovered about 3x more stars within 25\,arcsec from the centre of M92 compared to the temporally averaged image, with an average reduction in D50 measurements by a factor of 2.5 and an average 3.5x improvement in Strehl ratio. However, the performance metrics, particularly for fainter stars, indicated a more rapid decline in the real data scenario than in the simulated one. This highlights the challenges posed by real-world conditions, such as the intermittent turbulent ground layer observed in the M92 video stream and the non-Gaussian behaviour of pixel noise in sCMOS devices. The model's tendency to ``smooth" and/or possibly ``hallucinate" fainter stars is an expected limitation that warrants further investigation. The performance on real data, while promising, highlights areas of improvement, particularly in handling effects not currently accounted for in simulations. Indeed, given the suite of current limitations in the simulations mentioned in Section \ref{section:limitations}, it was not granted that meaningful inferences could be made on real data. 

In conclusion, this first ``DanceCam'' study presents a machine learning-based approach to addressing the challenge of atmospheric turbulence in astronomical imaging. The results obtained from both simulated and real data demonstrate the capabilities and potential of this method. However, it is important to acknowledge that further development and refinement are necessary for the DanceCam approach, particularly in enhancing the reconstruction of fainter stars and improving astrometric precision. The method, in its current form, shows promise for contributing to the field of astronomical research. Approaches like ours, while still in their nascent stages, could play a role in improving the clarity and accuracy of wide-field ground-based observations. Nevertheless, we recognize the need for cautious optimism regarding the method's scientific potential, and we encourage continued exploration and testing to fully ascertain its efficacy in diverse observational scenarios.


%% file: Sections/Appendix/appendix.tex
\newpage
\appendix

\section{Comparison to inference on temporally averaged frame}
\label{section:avgframeinference}

Our primary methodology for mitigating atmospheric turbulence in astronomical images utilizes short video sequences to train a Residual U-Net. An alternative approach involves training the U-Net with a single, long exposure image. This section reiterates the rationale behind our video-based method and the potential limitations associated with a long exposure input.

Atmospheric turbulence introduces dynamic, time-varying distortions in stellar light. A video stream, comprised of numerous short-exposure frames, effectively captures these temporal variations within the turbulence. By providing the U-Net with this sequence of frames, we enable it to analyze the patterns of turbulence and discriminate between photons belonging to different stars, even in cases where blurring leads to image overlap in single frames.

When instead presented with a long exposure image, the U-Net receives a composite view where the effects of turbulence are accumulated over time. This obscures the finer temporal patterns of turbulence, making it more difficult for the model to disentangle light from closely spaced astronomical sources.

\begin{figure*}
    \centering
    \includegraphics[width=1\textwidth]{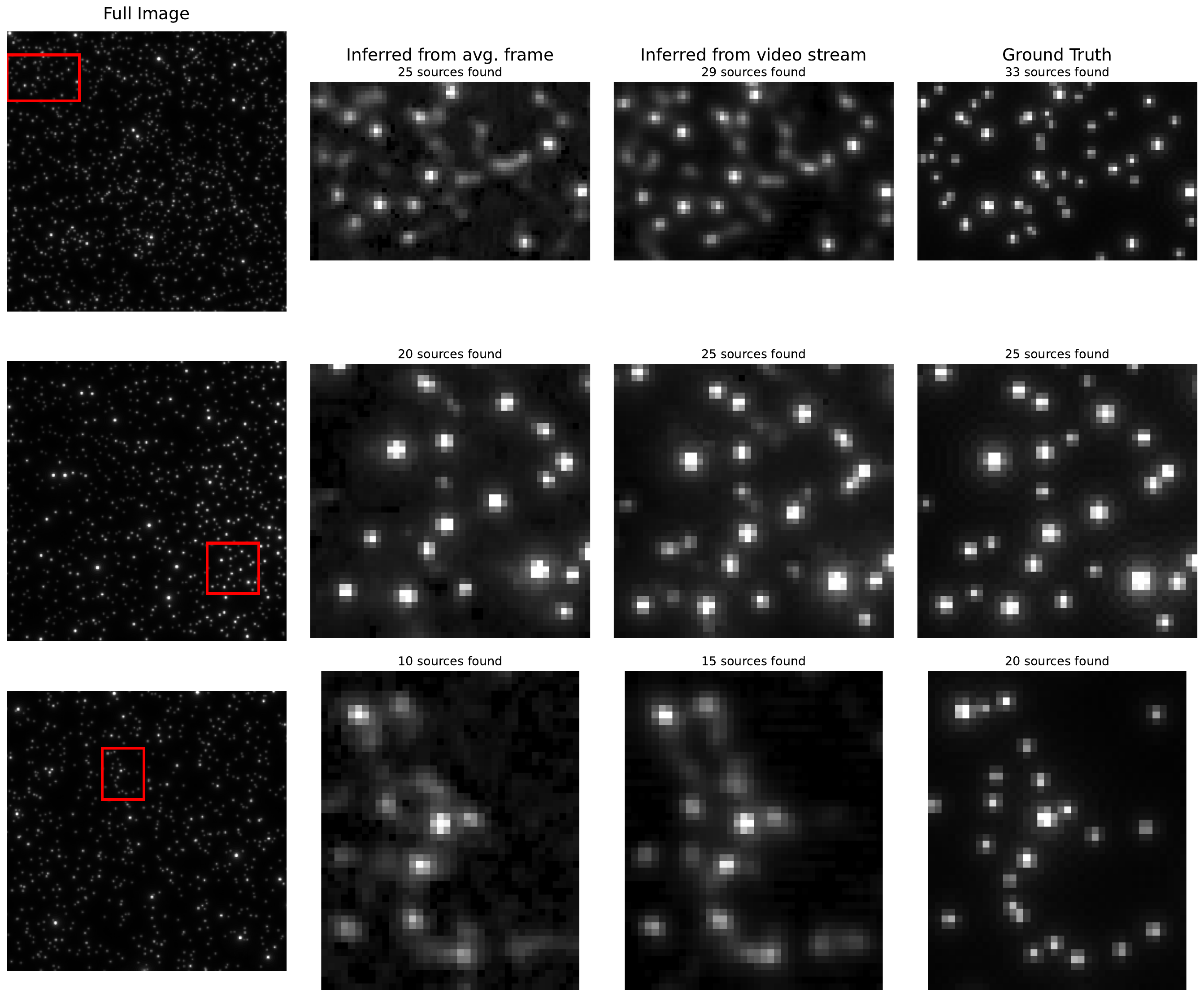}
    \caption{Qualitative performance comparison of source detection in crowded regions of simulated stellar fields. \texttt{SExtractor} is applied to the U-Net's inferred output when trained on either a video stream or its temporally averaged equivalent. The video-based approach demonstrates superior ability to discriminate light from nearby sources, as seen in both the source counts and the zoomed in densely-crowded sub-regions (indicated by red rectangles).}
    \label{fig:randomfields_singleframe_vs_multiframe_inference}
\end{figure*}

\begin{figure}
    \centering
    \includegraphics[width=0.5\textwidth]{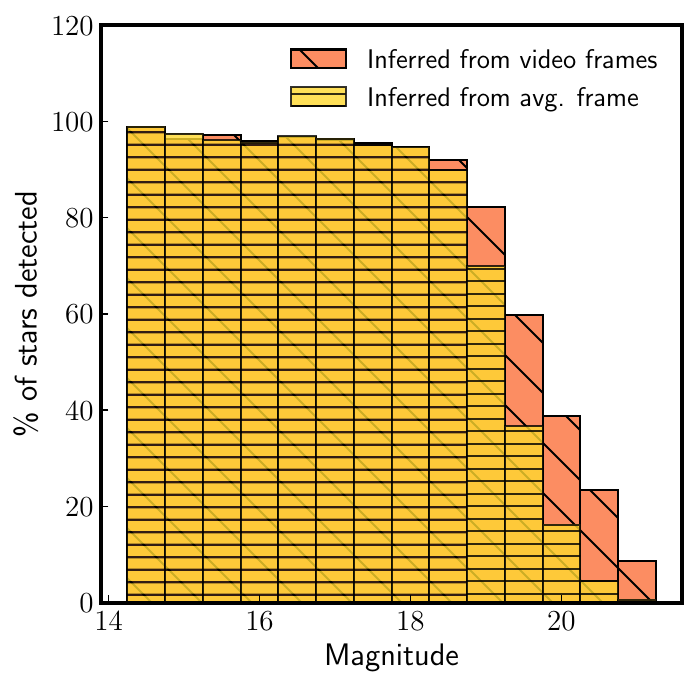}
    \caption{Quantitative performance comparison of source detection in crowded regions of simulated stellar fields. The same process as described in Figure \ref{fig:randomfields_singleframe_vs_multiframe_inference} was performed for several hundred simulated stellar fields and the number of detected stars compared to the ground truth was binned according to magnitude. The video-based approach again demonstrates a consistently better ability to discriminate light from neighbouring, especially faint sources.}
\label{fig:randomfields_singleframe_vs_multiframe_inference_percentstars}
\end{figure}

For comparative testing, we simulate long exposure images by temporally averaging the frames within a video stream. This averaging process mimics the effect of a traditional long exposure observation. Figure \ref{fig:randomfields_singleframe_vs_multiframe_inference} shows a \textit{qualitative} comparison of sources detected by \texttt{SExtractor} -- in particularly crowded regions of randomly generated stellar fields -- between output images inferred from either a video stream or a single temporally averaged frame. For a more \textit{quantitative} analysis, we test on several hundred simulated stellar fields to collect statistics of the detected sources in crowded sub-regions. As seen in Figure \ref{fig:randomfields_singleframe_vs_multiframe_inference_percentstars}, inferring on video frames allows for recovery of $\sim$2x more stars in the fainter end of the magnitude distribution (m $>$ 19). These simple tests help to confirm that the video streams enable the U-Net to better disentangle the light from neighbouring stars and consistently recover more sources in crowded regions. 



While the long exposure input approach may provide practical benefits, such as a reduced computational cost and a simplified data acquisition process, our research demonstrates that the video-based method holds superior potential for reconstructing turbulence-free astronomical images. By exploiting the temporal dynamics of atmospheric distortions, our approach enables the Residual U-Net to more effectively reconstruct stellar images, ultimately enhancing the resolution of astronomical images.

\section{Results with a smaller temporal input context}
\label{section:appendix_smaller_context}

To investigate the impact of the input temporal context on U-Net training and testing, we conducted experiments with a reduced context window of 6 seconds (32 frames @ 5.25 frames per second). This appendix presents the results of these experiments, offering a comparison to the performance achieved with a 12-second context.  We repeated the M92 tests outlined in Section \ref{section:testcaseM92}, with results presented below.

For simulated M92 observations, Figure \ref{fig:M92sim_percentstarsrecov_6s} shows the proportion of stars recovered across different magnitudes. While a 6-second context still achieves a respectable 85.8\% recovery rate and 96.4\% precision, the 12-second context (Figure \ref{fig:M92sim_percentstarsrecov}) demonstrates a noticeable advantage (86.2\% recovery rate and 97.8\% precision), especially for fainter stars.

\begin{figure*}
\centering
\includegraphics[width=0.9\textwidth]{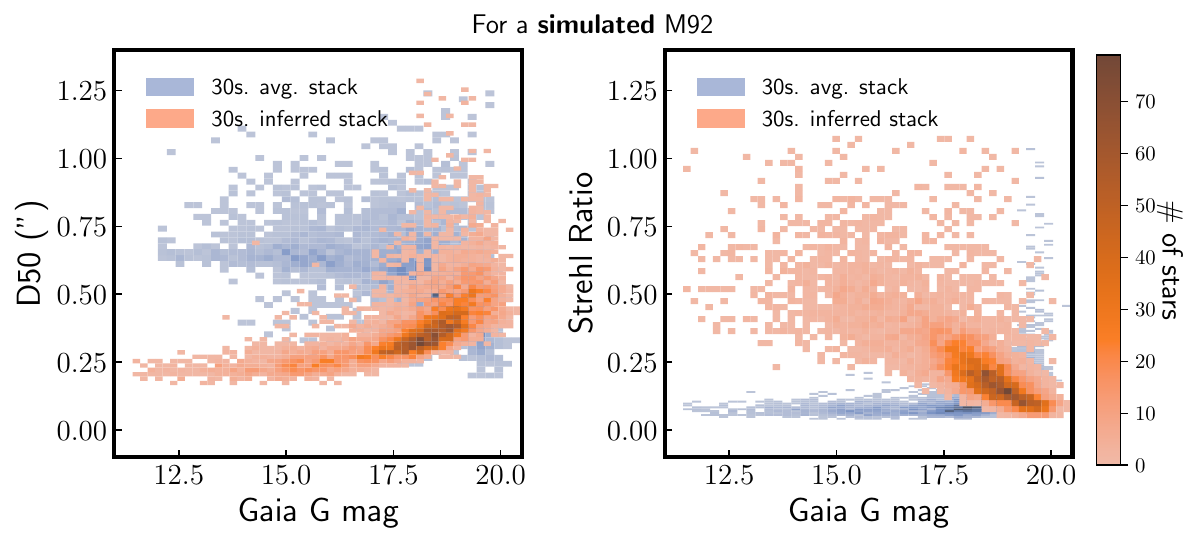}
\caption{D50 and Strehl ratio performance versus magnitude for simulated M92 observations given a 6-second temporal input to the U-Net (only relevant to the inferred stack). Compare to Figure \ref{fig:M92sim_d50_strehl} for the 12-second temporal context.}
\label{fig:M92sim_d50_strehl_6s}
\end{figure*}

Figure \ref{fig:M92sim_d50_strehl_6s} highlights performance differences for D50 and Strehl ratio measurements when compared to \ref{fig:M92sim_d50_strehl}. The 12-second context yields superior results (0.23" D50, 0.60 average Strehl ratio) compared to the 6-second context (0.32" D50, 0.51 Strehl ratio). Again, these benefits are most pronounced for fainter sources.

\begin{figure}
\centering
\includegraphics[width=0.5\textwidth]{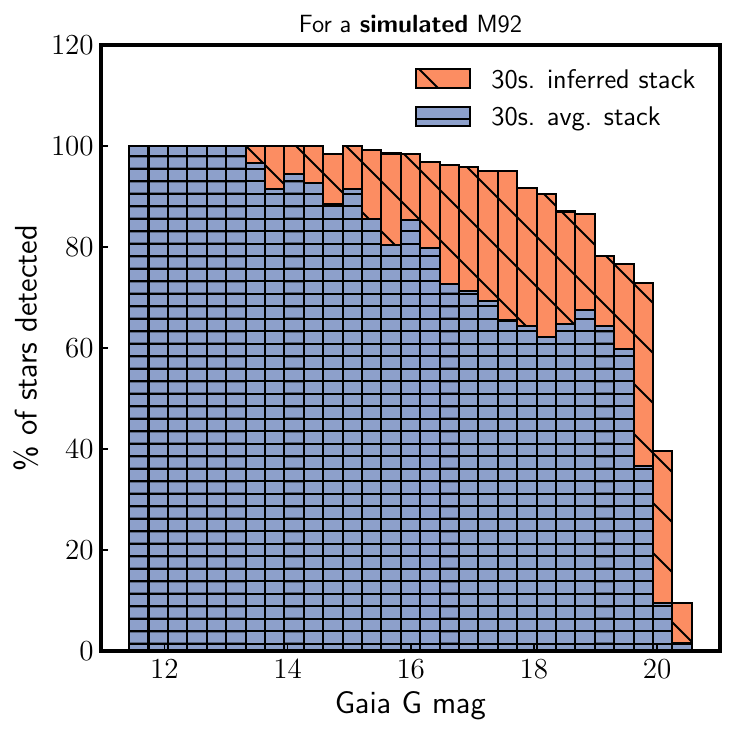}
\caption{Percentage of stars recovered by magnitude range for simulated M92 observations given a 6-second temporal input to the U-Net (only relevant to the inferred stack). Compare to Figure \ref{fig:M92sim_percentstarsrecov} for the 12-second temporal context.}
\label{fig:M92sim_percentstarsrecov_6s}
\end{figure}

